\documentclass[12pt,preprint]{aastex}
\slugcomment{Draft writing}
\shorttitle{ENERGY AND MOMENTUM TRANSFER VIA COULOMB FRICTIONS}
\shortauthors{ASANO et al.}

\begin{document}

\title{ENERGY AND MOMENTUM TRANSFER VIA COULOMB FRICTIONS
IN RELATIVISTIC TWO FLUIDS}
\author{\scshape K. Asano}
\affil{Division of Theoretical Astronomy, National Astronomical Observatory of Japan}
\email{asano@th.nao.ac.jp}
\and
\author{\scshape S. Iwamoto}
\affil{Software Cradle Co. Ltd.}
\and
\author{\scshape F. Takahara}
\affil{Department of Earth and Space Science, Graduate School of Science, Osaka University}
\email{takahara@vega.ess.sci.osaka-u.ac.jp}

\date{Submitted; accepted }

\begin{abstract}

We numerically calculate the energy and momentum transfer rates
due to Coulomb scattering
between two fluids moving with a relative velocity.
The results are fitted by simple functions.
The fitting formulae are useful to simulate
outflows from active galactic nuclei and compact high energy sources.

\end{abstract}


\section{INTRODUCTION}
\label{INTRODUCTION}

Relativistic jets are observed in active galactic nuclei (AGN) and
Galactic black hole candidates.
The velocity of these jets are highly relativistic
with a bulk Lorentz factor above 10.
The kinetic power is almost comparable to the
Eddington luminosity.
The production and bulk acceleration of these jets
are still unknown, though many ideas have been
proposed ranging from magneto-hydrodynamical to
radiative and thermal ones.
There is no consensus on how jets are produced and
accelerated.

Although it is difficult to determine the matter content of jets
from observations, several independent arguments favor electron-positron jets
\citep{tak94,tak97,rey96,war98,hom99,hir99,hir00,hir05,kin04,cro05}.
Electron-positron jets are most likely produced
in accretion disks around the central black holes.
Because the electron mass is much smaller than the proton mass,
the produced electron-positron pairs can be ejected more easily
than protons.
Some papers discuss the accretion disks
with electron-positron outflows \citep{mis95,lia95,li96,yam99}.
If the accretion disks form hot pair plasma
strongly coupled with photons,
the plasma may be thermally accelerated like the fireball
applied to gamma-ray bursts.
\citet{iwa02,iwa04} showed that a ``Wien fireball'',
which is optically thick to Compton scattering but thin
to absorption, results in a relativistic outflow
avoiding the difficulties of pair annihilation
and radiation drag.

Pairs are formed via photon-photon collisions
in the accretion disk composed of normal plasma (electron-proton).
Such pairs may escape from the disk by their own pressure
or radiative force \citep{yam99}.
In order to investigate pair ejection from the disk,
one needs to treat multi-component plasma dynamics with radiation field.
However, there has been no study of formation and ejection of
pairs from the accretion disk taking into account radiative
transfer consistently.
Such studies require knowledge of friction force between
the background plasma and pair outflow.
Although several plasma effects may be important,
Coulomb scattering is the first one to be taken into account.
The purpose of this paper is to obtain useful formulae
of the energy and momentum transfers between two fluids
moving with a relative velocity via Coulomb interaction.

The heating and cooling rates of thermal plasmas are
well known in both non-relativistic \citep{spi56}
and relativistic cases \citep{ste83a}.
\citet{der85} analytically derived the energy exchange rate of
two isotropic plasmas interacting at different temperatures.
If there exists a relative velocity between two plasmas,
analytical expressions of the energy and momentum exchange rates
are hard to be obtained.

In this paper, we numerically obtain the energy and momentum exchange rates
due to Coulomb scattering
between two Maxwell-Boltzmann plasmas with a relative velocity.
In another paper, we simulate pair outflows from hot plasmas
\citep{asa06} using the results in this paper.
In \S 2 we describe a formulation of the reaction rates between
two plasmas.
In \S 3 we show the numerical results for parameter regions
we are interested in.
Fitting formulae are presented so that one can use
our results in a numerical simulation of two fluid dynamics.

\section{TWO-FLUIDS INTERACTION}
\label{TWO-FLUIDS INTERACTION}

In this section we describe a formulation
of the exchange rates of energy and momentum due to
Coulomb scattering in two fluids.
Consider a collision of two particles, which belong to the fluids A and B,
respectively.
Before the collision their 4-momenta in the laboratory frame are 
\begin{equation}
p_{\rm A}^\mu = m_{\rm A}  c \gamma_{\rm A} (1,\mbox{\boldmath $\beta$}_{\rm A}),
\quad p_{\rm B}^\mu = m_{\rm B}  c \gamma_{\rm B} (1,\mbox{\boldmath $\beta$}_{\rm B}). 
\end{equation}
Here, $m_{\rm i}$ and $\mbox{\boldmath $\beta$}_{\rm i}$ are the mass and
velocity normalized by the light speed $c$ of the particle in the fluid i,
respectively.
The Lorentz factor of the particle is denoted as
$\gamma_{\rm i}=(1-\beta_{\rm i}^2)^{-1/2}$, and
that of the relative velocity $\beta_{\rm r}$ is written
as
\begin{equation}
\gamma_{\rm r}=\gamma_{\rm A} \gamma_{\rm B}
(1- \mbox{\boldmath $\beta$}_{\rm A} \cdot \mbox{\boldmath $\beta$}_{\rm B}),
\end{equation}
which is Lorentz invariant.

The velocity of the center-of-mass (CM) in the laboratory frame is 
\begin{equation}
\mbox{\boldmath $\beta$}_{\rm CM} = 
{m_{\rm A} \gamma_{\rm A} \mbox{\boldmath $\beta$}_{\rm A}
+ m_{\rm B} \gamma_{\rm B} \mbox{\boldmath $\beta$}_{\rm B} \over
m_{\rm A} \gamma_{\rm A} + m_{\rm B} \gamma_{\rm B}}.
\end{equation}
Let this direction define the $x$-axis in the Cartesian coordinate
$(x,y,z)$ in the laboratory frame, $\mbox{\boldmath $\beta$}_{\rm CM}=
\beta_{\rm CM} \hat{x}$. 
The velocity of particles in the fluid i is decomposed as
\begin{equation}
\begin{array}{rcl}
\mbox{\boldmath $\beta$}_{\rm i} &=& \beta_{{\rm i}x} \hat{x} +
\beta_{{\rm i}y} \hat{y} + \beta_{{\rm i}z} \hat{z},
\end{array}
\end{equation}

Denoting the 4-momentum of particles in the fluid A
after the collision as $p_{\rm A2}^\mu= m_{\rm A}c
\gamma_{\rm A2}(1, \mbox{\boldmath $\beta$}_{\rm A2})$,
the exchange of 4-momentum is written as
\begin{equation}
\Delta p_{\rm A}^\mu= m_{\rm A} c (\gamma_{\rm A2}-\gamma_{\rm A},\gamma_{\rm A2}
\mbox{\boldmath $\beta$}_{\rm A2}-\gamma_{\rm A} \mbox{\boldmath $\beta$}_{\rm A}). 
\end{equation}
The mean exchange of energy for elastic scattering in the laboratory frame is 
\begin{eqnarray}
\langle \Delta E \rangle & = &
\langle \Delta p_{\rm A}^0 \rangle c  
=  - \langle \Delta p_{\rm B}^0 \rangle c  \nonumber \\
 & = & 2 m_{\rm A} c^2
\gamma_{\rm A} \gamma_{\rm CM}^2 \sin^2(\alpha/2) \ \beta_{\rm CM}  
(\beta_{\rm CM} 
- \beta_{{\rm A}x})  \nonumber \\
 & = & {-2 m_{\rm A} m_{\rm B} \sin^2 (\alpha/2)  \over m_{\rm A}^2 
+ m_{\rm B}^2 
+ 2 \gamma_{\rm r} m_{\rm A} m_{\rm B} } c^2   \label{de} \\
& & \times \left[ m_{\rm B} \gamma_{\rm A} - m_{\rm A} \gamma_{\rm B} 
+ \gamma_{\rm r} (m_{\rm A}\gamma_{\rm A}- m_{\rm B}\gamma_{\rm B}) \right]. 
\nonumber
\end{eqnarray}
\citep{ste83a}, where
$\alpha$ is the scattering angle in the CM-frame and
$\gamma_{\rm CM}=(1-\beta_{\rm CM}^2)^{-1/2}$.
The mean momentum exchange in the laboratory frame is described as 
\begin{eqnarray}
\langle \Delta p_{\rm A}^x \rangle & = & 
 2 m_{\rm A} c \gamma_{\rm A} \gamma_{\rm CM}^2 (\beta_{\rm CM} - \beta_{{\rm A}x})  \sin^2 (\alpha/2)  \nonumber  \\
\langle \Delta p_{\rm A}^y \rangle & = & 
 -2 m_{\rm A} c \gamma_{\rm A} \beta_{{\rm A}y} \sin^2 (\alpha/2) \label{dp} \\
\langle \Delta p_{\rm A}^z \rangle & = & 
 -2 m_{\rm A} c \gamma_{\rm A} \beta_{{\rm A}z} \sin^2 (\alpha/2)  \nonumber
\end{eqnarray}

Hereafter we denote
the distribution functions of the fluid i
as $n_{\rm i} f_{\rm i}(\mbox{\boldmath $p$}_{\rm i})$,
where $n_{\rm i}$ is the number density of the fluid
in the laboratory frame
so that $\int f_{\rm i}(\mbox{\boldmath $p$}_{\rm i}) d^3 p_{\rm i}=1$.
The scattering rate per unit volume,
\begin{equation}
R_{\rm AB}=n_{\rm A} f_{\rm A}(\mbox{\boldmath $p$}_{\rm A})
n_{\rm B} f_{\rm B}(\mbox{\boldmath $p$}_{\rm B})
c {\gamma_{\rm r} \beta_{\rm r} \over \gamma_{\rm A} \gamma_{\rm B}}
\frac{d \sigma(\mbox{\boldmath $p$}_{\rm A},\mbox{\boldmath $p$}_{\rm B})}
{d \Omega}
d \Omega,
\end{equation}
\citep{wea76,lan75}, is Lorentz-invariant.
In this expression $d\Omega \equiv 2 \pi \sin{\alpha} d \alpha$
and $d \sigma/d \Omega$ is the differential cross-section.
The exchange rate of 4-momentum per unit volume $dV$ and per unit time $dt$
is described as 
\begin{equation}
{dP^\mu \over dt dV} = {1 \over 1+\delta_{\rm AB}}
\int \int \langle \Delta p^\mu  \rangle R_{\rm AB}
d^3 p_{\rm A} d^3 p_{\rm B}, 
\label{exc}
\end{equation}

Then, we now restrict the situation that
particles in the fluid A have an isotropic Maxwell-Boltzmann (MB) distribution
in the laboratory frame $K$ and that particles in the fluid B have another 
isotropic MB
distribution in another inertial frame $K'$.
The relative velocity between the two frames is defined as
$\beta_{\rm R}$ and its Lorentz factor is
$\Gamma_{\rm R}=(1-\beta_{\rm R}^2)^{-1/2}$.

The relativistic MB distribution of the fluid A in the laboratory frame
is described as
\begin{equation}
f_{\rm A}(\mbox{\boldmath $p$}_{\rm A}) d^3 p_{\rm A} =
{\exp(-\gamma_{\rm A}/\theta_{\rm A}) \over 4 \pi \theta_{\rm A} K_2(1/\theta_{\rm A})}
u_{\rm A}^2   d u_{\rm A} d \Omega_{\rm A},
\label{disa}
\end{equation}
where $u_{\rm i}=\gamma_{\rm i} \beta_{\rm i}$,
$\theta_{\rm i} \equiv T_{\rm i}/m_{\rm i} c^2$
is the temperature normalized by the mass of the particles,
$K_2$ is the modified Bessel function of the 2nd kind,
$d \Omega_{\rm i} \equiv d \mu_{\rm i} d \phi_{\rm i}$  is the solid angle of
$\mbox{\boldmath $p$}_{\rm i}$, respectively.
Because of Lorentz invariance of the distribution function,
$n_{\rm B} f_{\rm B}=n'_{\rm B} f'_{\rm B} =
n'_{\rm B} \exp(-\gamma'_{\rm B}/\theta_{\rm B})/
(4 \pi \theta_{\rm B} K_2(1/\theta_{\rm B}))$, 
where primed values represent quantities in the frame $K'$.
Since $n'_{\rm B}=n_{\rm B}/\Gamma_{\rm R}$, we obtain 
\begin{equation}
f_{\rm B}(\mbox{\boldmath $p$}_{\rm B}) d^3 p_{\rm B} =
{\exp (-\gamma'_{\rm B}/\theta_{\rm B}) \over 4 \pi \Gamma_{\rm R}
\theta_{\rm B} K_2(1/\theta_{\rm B})} u_{\rm B}^2
d u_{\rm B} d \Omega_{\rm B},
\label{disb}
\end{equation}
where $\gamma'_{\rm B}=\Gamma_{\rm R} (\gamma_{\rm B}- \beta_{\rm R} \mu_{\rm B}
u_{\rm B})$.

When the particle species in the fluids A and B are the same, 
we cannot determine if the scattered particle belongs to fluid A
or B by the quantum mechanics principle. In this case, we treat as 
follows.
As is well known, plasma relaxation is achieved
by mainly small angle scattering ($\alpha \ll 1$).
Therefore, it may be natural to consider that $\alpha \leq \pi/2$
in this case.
Even if we allow large angle scattering ($\alpha>\pi/2$),
we ought to consider that it means
exchange of particles between the fluids A and B.
As a result, we can use the above formulation assuming $\alpha \leq \pi/2$
for scatterings of the same species of particles.

The integral over $\alpha$ is analytically possible
(see Appendix).
On the other hand, from the axial symmetry
the sextuple integral in equation (\ref{exc})
is reduced to a quintuple integral.
However, further reduction in the order of integral may not be carried out
differently from isotropic cases such as in \citet{der85}.
Therefore, a straightforward method in numerical integration
is inefficient in this case.
Using a Monte Carlo technique that is similar to
the method in \citet{ram81}, we numerically integrate equation (\ref{exc})
with $N=10^9$ collisions of two particles whose momentum distributions
are proportional to equations (\ref{disa}) and (\ref{disb}), respectively.
The differential cross-sections we used
are summarized in Appendix.
The Coulomb logarithm $\ln{\Lambda}=20$ is adopted throughout this paper.

\section{NUMERICAL RESULTS}

We calculate energy and momentum transfers (ET and MT)
for probable temperatures of leptons and protons
in hot accretion disks.
The relative velocity of two fluids,
$U_{\rm R}=\sqrt{\Gamma_{\rm R}^2-1}$,
is in the range of $10^{-2} \leq U_{\rm R} \leq 10^2$
in our calculation.
Because of the property of the Monte Carlo method,
estimated values of ET and MT may
have errors especially for MT with $U_{\rm R} \ll 1$.
However, from our experiences, the errors are at most 10 \% 
for $u_{\rm R}>0.1$.
In contrast to the uncertainty in the Coulomb logarithm,
these errors  may be negligible.

We express the energy gain rate of the outflowing plasma (fluid B) as
\begin{equation}
{d E \over dt dV}=n_{\rm A} n_{\rm B} m_{\rm e} c^2 F_{\rm E}(T_{\rm A},
T_{\rm B},U_{\rm R}).
\end{equation}
in the laboratory frame (the comoving frame of the fluid A).
The value $F_{\rm E}$ has a dimension of $[\mbox{cm}^3/\mbox{s}]$.
We should notice that $n_{\rm A}$ and $n_{\rm B}=\Gamma_{\rm R} n'_{\rm B}$
are densities in the laboratory frame.

On the other hand, the momentum loss rate of the outflowing plasma is expressed as
\begin{equation}
{d P \over dt dV}=n_{\rm A} n_{\rm B} m_{\rm e} c F_{\rm P}(T_{\rm A},
T_{\rm B},U_{\rm R}).
\end{equation}
The dimension of $F_{\rm P}$ is the same as that of $F_{\rm E}$.

Since $dt dV$ is Lorentz invariant,
these rates in the laboratory frame correlate with
the rates in the comoving frame of the fluid B as
\begin{equation}
{d E \over dt dV}=\Gamma_{\rm R} {d E' \over dt' dV'}-c U_{\rm R} {d P' \over dt' dV'}
\end{equation}
\begin{equation}
{d P \over dt dV}=\Gamma_{\rm R} {d P' \over dt' dV'}-
\frac{U_{\rm R}}{c} {d E' \over dt' dV'},
\end{equation}
respectively, where the minus signs in the second terms are due to
the definition of $dP/dt dV$ (momentum ``loss'' rate).
The above equations show that negative energy gain rate (${d E/dt dV}<0)$
does not necessarily imply a temperature drop of the fluid B.
If the fluid B is decelerated rapidly ($d E'/dt'dV' \ll c d P'/dt' dV'$),
the energy gain rate in the laboratory frame can be negative.

\subsection{ENERGY TRANSFER in p-e INTERACTION}

In this subsection we show the numerical results of
ET between the background protons (A$=$p)
and outflowing electrons or positrons (B$=$e).
We assume the temperature of protons
is higher than that of the electrons (positrons).
Hereafter we denote the temperature of protons
normalized by electron mass as $\Theta_{\rm p} \equiv T_{\rm p}/
m_{\rm e} c^2=(m_{\rm p}/m_{\rm e}) \theta_{\rm p}$.
In Figure 1 we plot $F_{\rm E}$ for
$\Theta_{\rm p}=10$.
If $U_{\rm R}=0$, the energy gain rate is analytically obtained as
\begin{eqnarray}
F_{\rm E}(T_{\rm p}, T_{\rm e},0)=\frac{3 \sigma_{\rm T} }{2 m_{\rm p} c}
\frac{T_{\rm p}-T_{\rm e}}
{K_2(1/\theta_{\rm e}) K_2(1/\theta_{\rm p})} \ln{\Lambda} \times \nonumber \\
\left[
\frac{2 (\theta_{\rm e}+\theta_{\rm p})^2+1}{\theta_{\rm e}+\theta_{\rm p}}
K_1 \left(\frac{\theta_{\rm e}+\theta_{\rm p}}{\theta_{\rm e} \theta_{\rm p}}\right)
+2 K_0 \left( \frac{\theta_{\rm e}+\theta_{\rm p}}{\theta_{\rm e} \theta_{\rm p}}\right)
\right],
\end{eqnarray}
\citep{ste83b}.
Our numerical results for $U_{\rm R} \ll 1$ agree with the above analytical estimate.
For $U_{\rm R} \gtrsim 1.0$, $F_{\rm E}$ declines with $U_{\rm R}$.
If $U_{\rm R}$ is substantially larger than $\theta_{\rm e}$,
we can neglect the thermal motion of leptons and $F_{\rm E}$
becomes independent of $\theta_{\rm e}$.
Therefore, $F_{\rm E}$ agree with each other for $U_{\rm R} \gg \theta_{\rm e}$.

\begin{figure}[t]
\centering
\epsscale{1.0}
\plotone{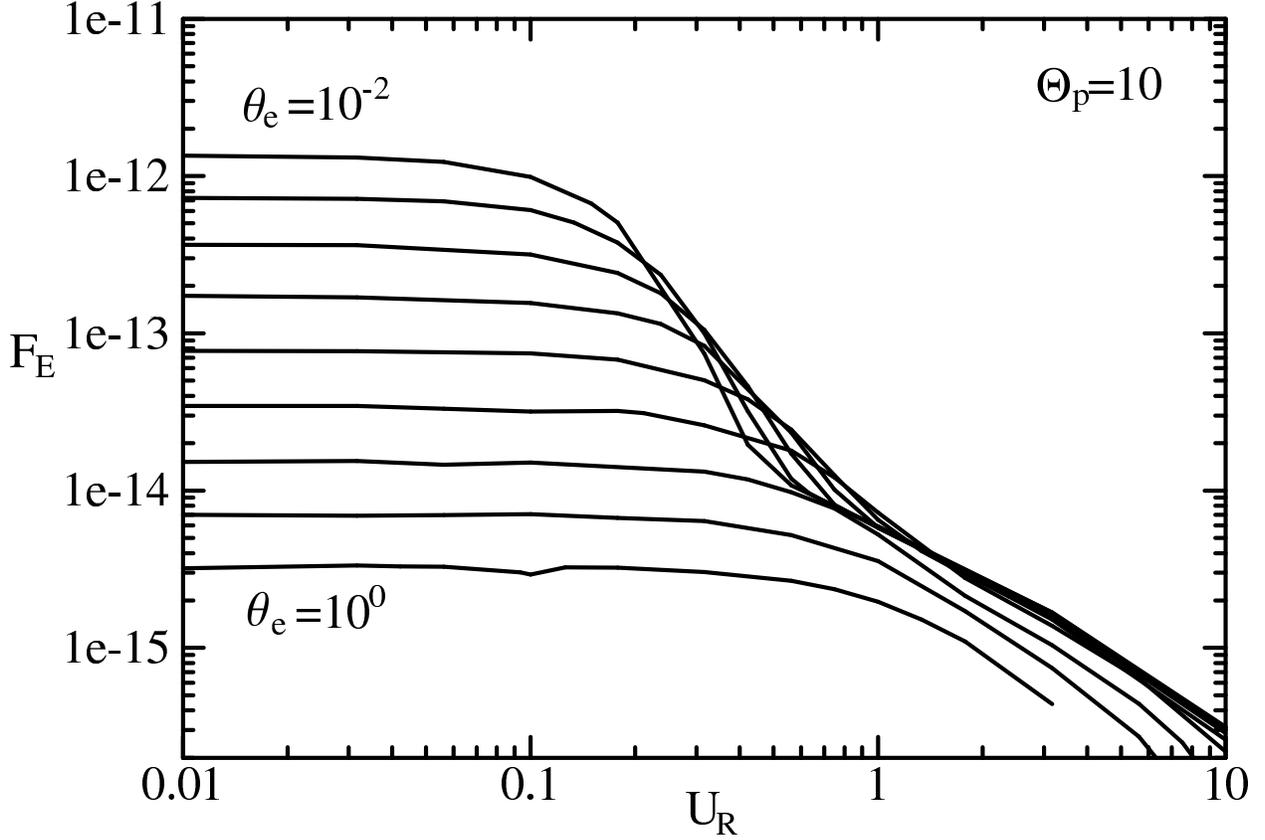}
\caption{
$F_{\rm E}$ of p-e interaction for $\Theta_{\rm p}=10.0$
in cgs unit. The temperatures of electrons (positrons), $\theta_{\rm e}$,
are from $10^{-2}$ to $10^0$ at $10^{0.25}$
intervals in logarithmic scale.
}
\end{figure}

\begin{figure}
\centering
\epsscale{1.0}
\plotone{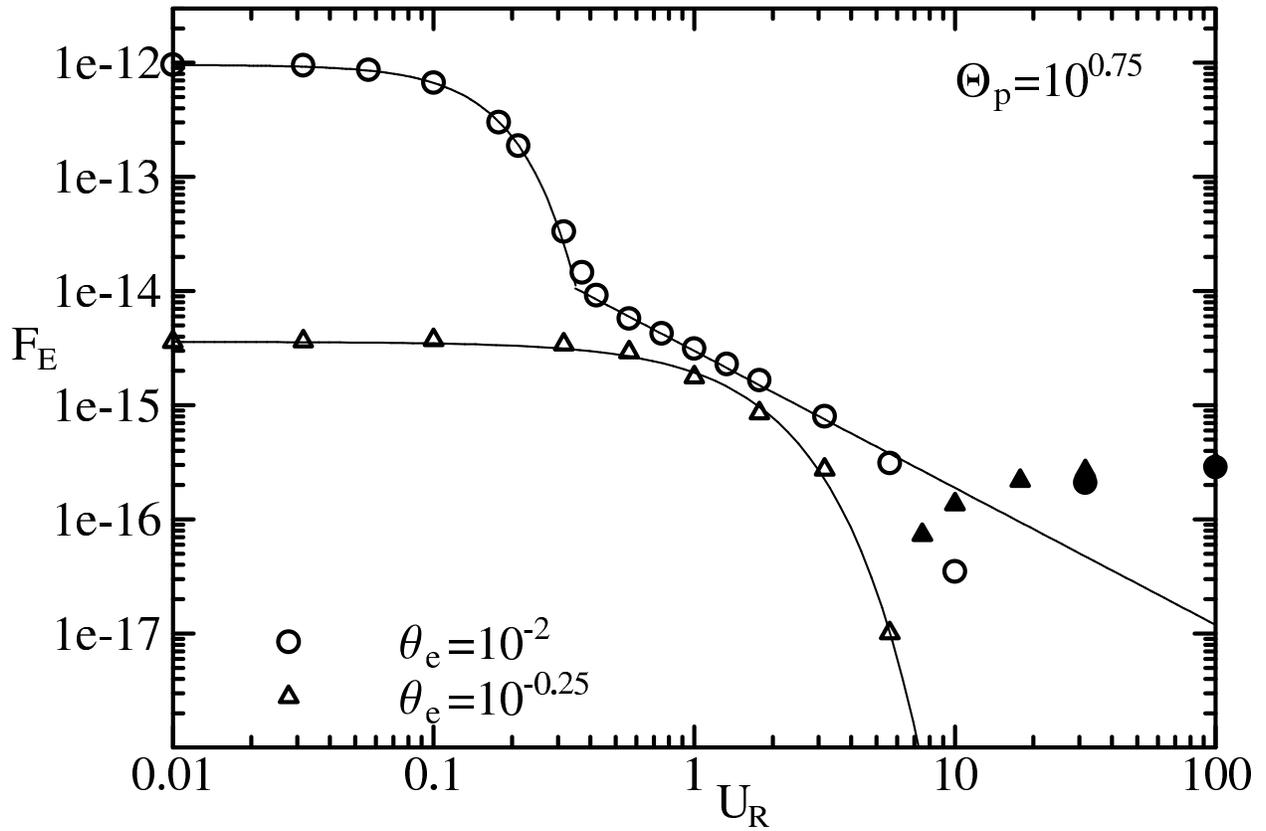}
\caption{
Two examples of the fitting functions for $F_{\rm E}$ of p-e interaction.
Circles and triangles are numerical results.
Filled symbols are $|F_{\rm E}|$ for negative $F_{\rm E}$.
Solid lines are the fitting functions we obtained
(see equation (\ref{fepe}) and Table 1).
}
\end{figure}

\begin{figure}
\centering
\epsscale{1.0}
\plotone{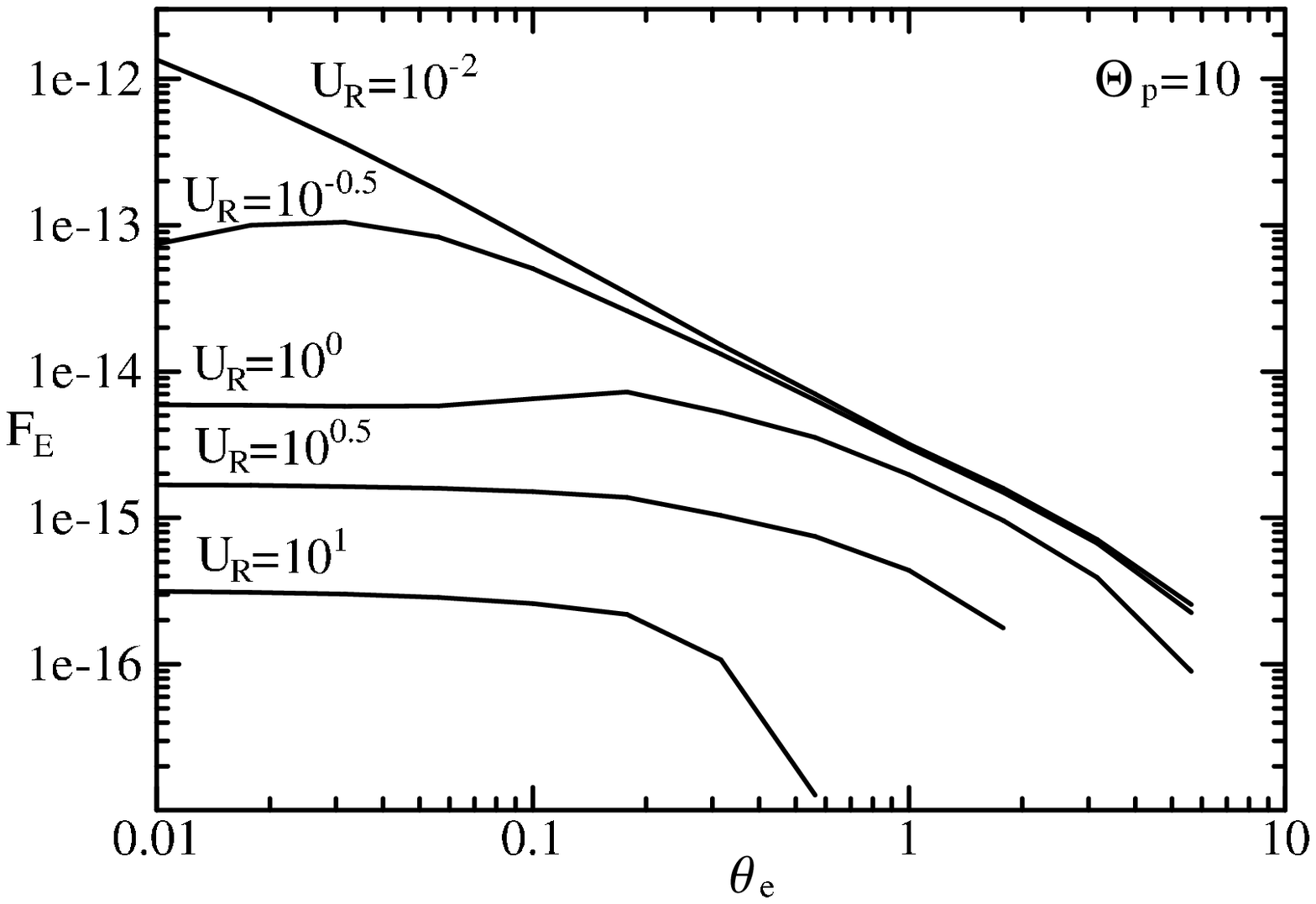}
\caption{
Same as Fig.1 but plotted vs. $\theta_{\rm e}$.
}
\end{figure}

Unfortunately, we could not fit our results by a simple analytical
function of $\Theta_{\rm p}$, $\theta_{\rm e}$, and $U_{\rm R}$.
In Figure 2 we plot $F_{\rm E}$ for $\Theta_{\rm p}=10^{0.75}$ 
as well as fitting functions. We fit the numerical results by two functions 
of $U_{\rm R}$; exponentially damped one for lower $U_{\rm R}$,
and power-law one for higher $U_{\rm R}$ as
\begin{eqnarray}
F_{\rm E}=\left\{
\begin{array}{ll}
A_{\rm L} \exp{\left[-\left( \frac{U_{\rm R}}{U_0} \right)^p \right]},
& \quad \mbox {for $U_{\rm R} < U_{\rm th}$}\\
A_{\rm H} U_{\rm R}{}^r, & \quad  \mbox {for $U_{\rm R} \geq U_{\rm th}$}.
\end{array} \right.
\label{fepe}
\end{eqnarray}
The energy gain rate of leptons $F_{\rm E}$
generally decreases 
with $U_{\rm R}$ and for $U_{\rm R} \gg \Theta_{\rm p}$
it becomes negative [see eq. (\ref{de})]. 
This corresponds to the deceleration of the lepton beam moving
through cold medium. 
Since $U_{\rm R} \ll m_{\rm p}/m_{\rm e}$ in our case,
$\beta_{\rm CM} \ll 1$, namely the CM frame is almost the
same as the laboratory frame.
For $U_{\rm R} \gg \Theta_{\rm p}$,
the Lorentz factor of electrons
in the CM frame $\gamma$ (see APPENDIX) is approximated as $\sim \gamma_{\rm e}
\sim U_{\rm R} \sim \gamma_{\rm r}$.
The equation (\ref{de}) indicates the energy loss of electrons in one collision
$\langle \Delta E \rangle \sim (m_{\rm e}/m_{\rm p}) U_{\rm R}^2 m_{\rm e} c^2
\sin^2{(\alpha/2)}$,
while the effective cross-section (see APPENDIX) $\sigma_{\rm eff}
\propto \gamma^{-2} \propto U_{\rm R}^{-2}$.
Therefore, $F_{\rm E}$ (roughly proportional to
$c \sigma_{\rm eff} \langle \Delta E \rangle$) becomes negative and constant
for $U_{\rm R} \gg \Theta_{\rm p}$ as shown in Figure 2.

Even for positive $F_{\rm E}$ the fitting functions (\ref{fepe}) deviate
for $U_{\rm R} \sim \Theta_{\rm p}$ as shown in Figure 2.
However, the fitting formulae may be practically correct
for $U_{\rm R} \lesssim \Theta_{\rm p}$.
The fitting parameters $A_{\rm L}$, $p$, $U_0$, $A_{\rm H}$, $r$, 
and $U_{\rm th}$
are listed in Tables 1-5.
The parameter $p$ is fixed as $2$ except for Tables 1 and 5.
We listed only for $\Theta_{\rm p}=10^{0.75}$, $10^1$, $10^{1.5}$, $10^{2}$,
and $10^{2.5}$.
One can interpolate $F_{\rm E}$ for general values of $\Theta_{\rm p}$
from our fitting formulae. Our fitting formulae give a satisfactory fit 
to numerical results for these ranges of $\Theta_{\rm p}$ and $U_{\rm R}$. 

Although the numerical results smoothly change with $\theta_{\rm e}$
as shown in Figure 1,
the fitting parameters do not necessarily change monotonically.
There exist numerical uncertainties of $\sim 10$ \% and  
the fitting method is not unique.
For example, if we refit the results with a slight change of the ``connecting 
point '' 
$U_{\rm th}$,
the other parameters, especially indices $p$, $q$ (see next subsection),
and $r$, will change substantially. 
Therefore, as will be shown throughout this paper,
parameter values we obtained do not always change systematically
with $\theta_{\rm e}$.

In Figure 3 we plot $F_{\rm E}$ for $\Theta_{\rm p}=10$
as functions of $\theta_{\rm e}$ for reference.
The results are not necessarily
monotonically decreasing functions of $\theta_{\rm e}$,
while curves in Figure 1 decrease with increasing $U_{\rm R}$.

\subsection{MOMENTUM TRANSFER in p-e INTERACTION}

\begin{figure}[t]
\centering
\epsscale{1.0}
\plotone{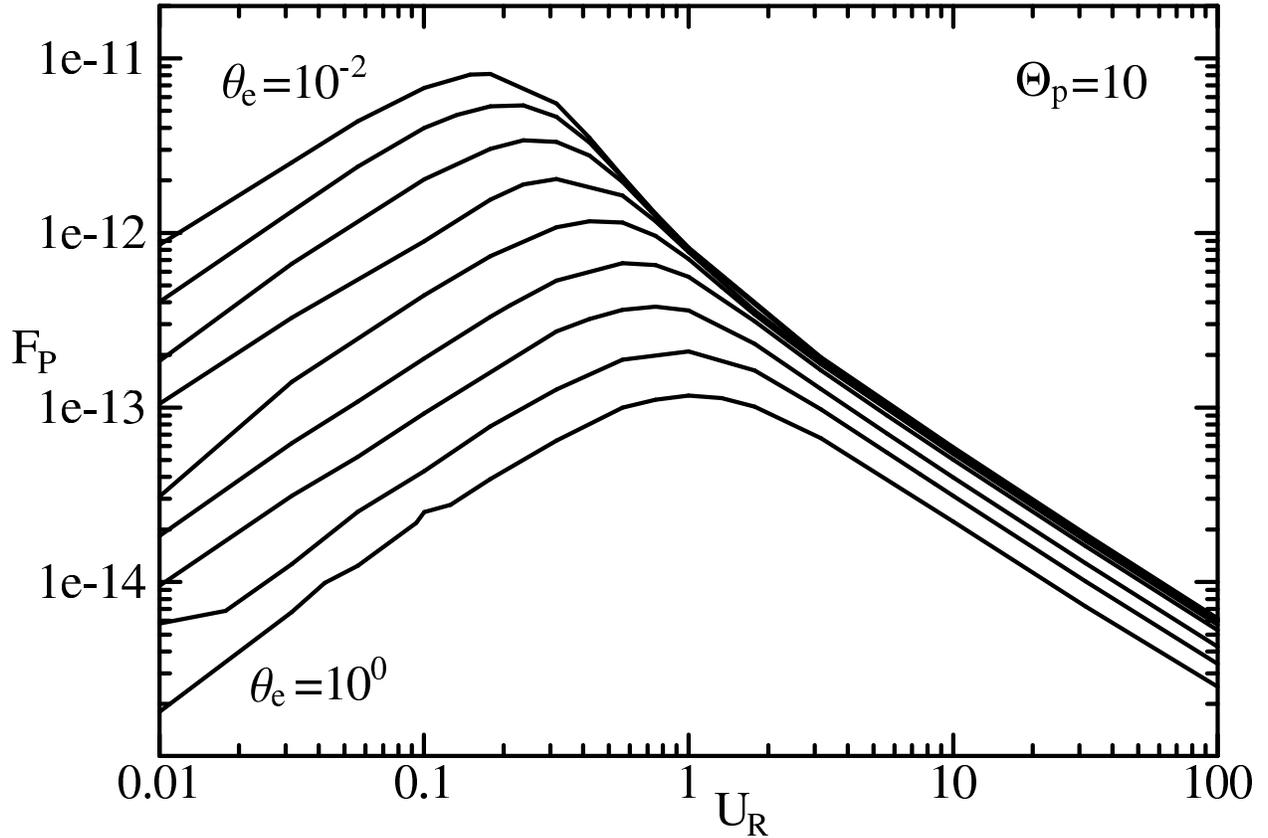}
\caption{
$F_{\rm P}$ of p-e interaction for $\Theta_{\rm p}=10.0$
in cgs unit. The temperatures of electrons (positrons)
are same as those in Fig. 1.
}
\end{figure}

\begin{figure}
\centering
\epsscale{1.0}
\plotone{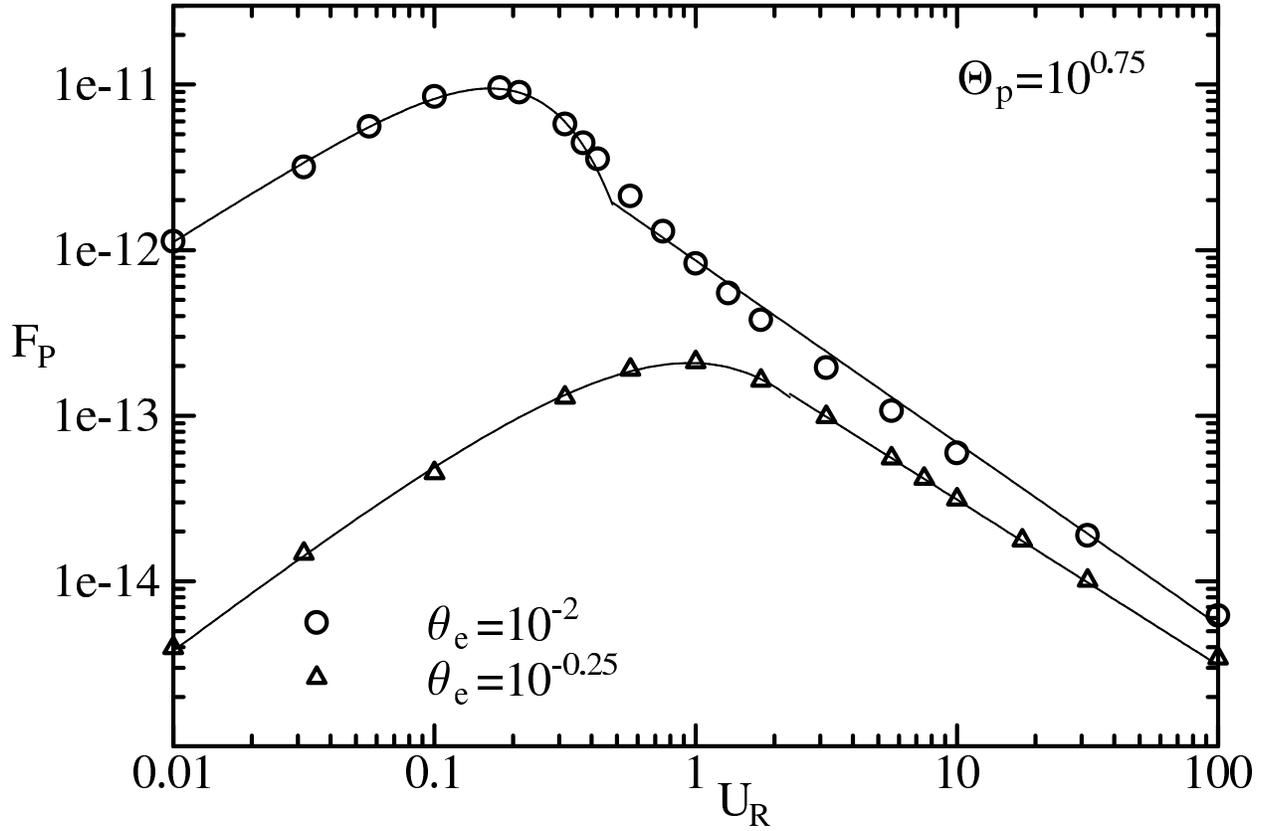}
\caption{
Two examples of the fitting functions for $F_{\rm P}$ of p-e interaction.
Circles and triangles are our numerical results.
Solid lines are the fitting functions we obtained
(see equation (\ref{fppe}) and Table 5).
}
\end{figure}

\begin{figure}
\centering
\epsscale{1.0}
\plotone{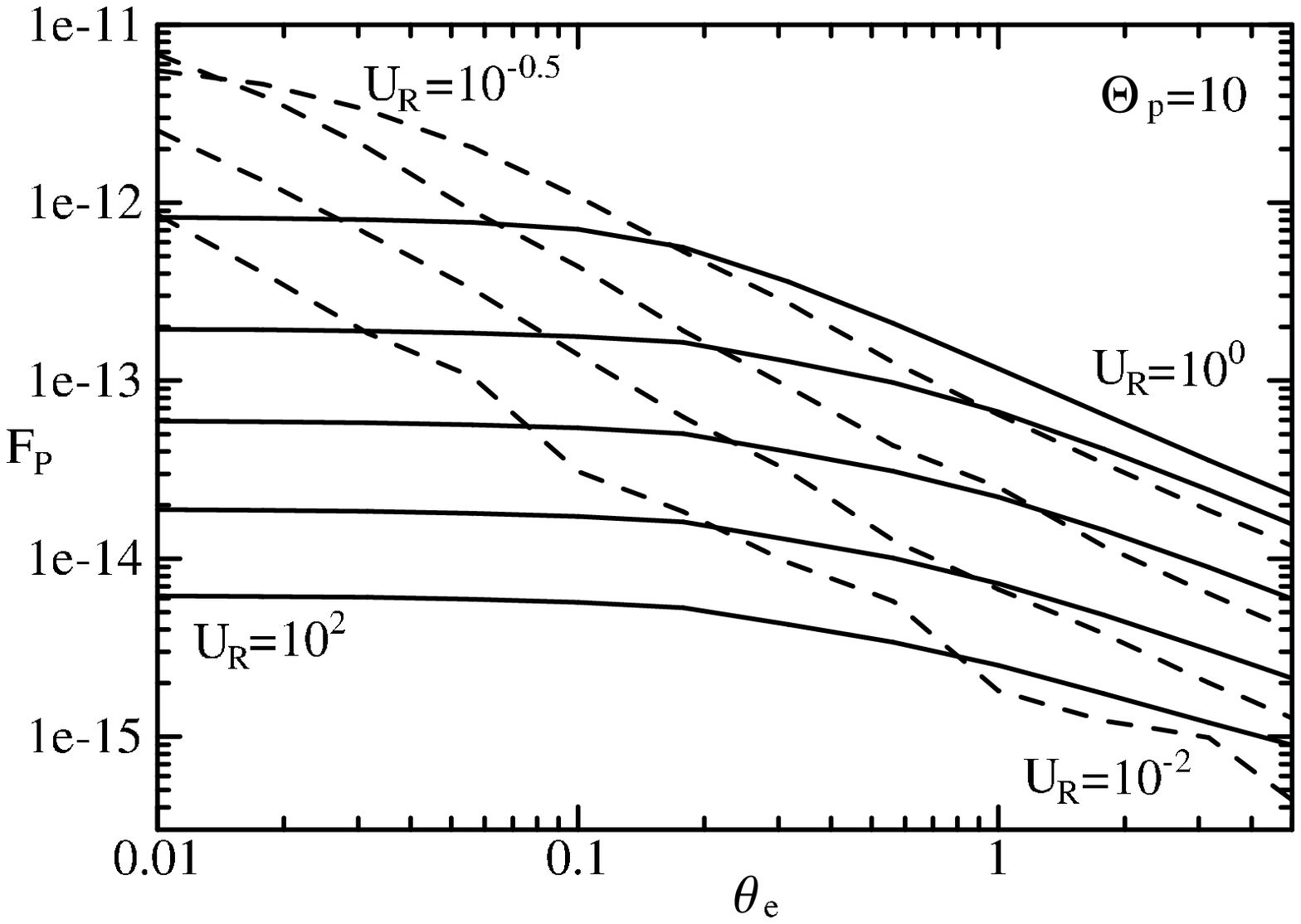}
\caption{
Same as Fig.4 but plotted vs. $\theta_{\rm e}$.
Dashed lines are from $U_{\rm R}=10^{-2}$ to $10^{-0.5}$,
and solid lines are from $U_{\rm R}=10^{0}$ to $10^{2}$
at $10^{0.5}$ intervals in logarithmic scale.
}
\end{figure}

The numerical results of $F_{\rm P}$ for $\Theta_{\rm p}=10$
are plotted in Figure 4.
The value of $F_{\rm P}$ increases with increasing $U_{\rm R}$
for $U_{\rm R} \ll 1$.
At $U_{\rm R} \sim 1$ $F_{\rm P}$ starts to decrease
because of smallness of the cross sections in relativistic collisions.
The results are fitted in the same way as $F_{\rm E}$.
The fitting formulae are written as
\begin{eqnarray}
F_{\rm P}=\left\{
\begin{array}{ll}
A_{\rm L} \exp{\left[-\left( \frac{U_{\rm R}}{U_0} \right)^p \right]}
\left( \frac{U_{\rm R}}{0.01} \right)^q,
& \mbox {for $U_{\rm R} < U_{\rm th}$},\\
A_{\rm H} U_{\rm R}{}^r, & \mbox {for $U_{\rm R} \geq U_{\rm th}$},
\end{array} \right.
\label{fppe}
\end{eqnarray}
and the parameters are listed in Tables 6-10.
As shown in Figure 5 the results are fitted by these functions
within $\sim 10$ \% errors for a wide range of $U_{\rm R}$.

Though a sufficient amount of momentum is exchanged for each collision of particles,
the total momentum exchange between the two fluids
is almost cancelled out for $U_{\rm R} \ll \theta_{\rm e}$.
Since $F_{\rm P}$ comes from the bulk motion of the outflowing electrons,
the behaviour, $F_{\rm P} \propto U_{\rm R}$ for $U_{\rm R} \ll 1$,
is a natural consequence.
On the other hand, $F_{\rm P}$ decreases with $U_{\rm R}$ for $U_{\rm R} \gtrsim 1$.
From equations (\ref{de}) and (\ref{dp}), the momentum loss of electrons
$\langle \Delta p \rangle \sim \langle \Delta E \rangle/(c \beta_{\rm CM})$
for $U_{\rm R} \gg \Theta_{\rm p}$.
As discussed in \S 3.1, $\langle \Delta E \rangle \propto U_{\rm R}^2$
and $\beta_{\rm CM} \propto U_{\rm R}$ in this case.
Then, we obtain $F_{\rm P} \propto U_{\rm R}^{-1}$, which is consistent
with the index $r$ in the Tables.

In Figure 6 we plot $F_{\rm P}$ for $\Theta_{\rm p}=10$
as functions of $\theta_{\rm e}$ for reference.
The line for $U_{\rm R}=10^{-2}$ is not smooth probably owing to
a statistical error in the Monte Carlo integral.
Although the rate of momentum transfer is
monotonically decreasing functions of $\theta_{\rm e}$,
the behavior beyond and below $U_{\rm R} \sim 1$ is apparently different. 
There exists a maximum value of $F_{\rm p}$ at $U_{\rm R}\sim 1$ 
and the Coulomb friction becomes small as $\theta_{\rm e}$ increases,
because of the monotonic behaviour of the cross section, $\sigma_{\rm eff}
\propto \gamma^{-2} \beta^{-4}$ (see APPENDIX).

\subsection{ENERGY TRANSFER in e-e INTERACTION}

In this subsection we show the results for ET between the background electrons
(A$=$be) and outflowing electrons (B$=-$) or positrons (B$=+$).
We assume that the temperature of the outflowing leptons is lower than
the background electrons.
In the opposite cases ET and MT are obtained from the Lorentz transformation
of our results.
The values for e${^-}$-e${^-}$ and e${^-}$-e${^+}$ interactions
are almost the same.
Therefore, we only show the results for e${^-}$-e${^-}$ hereafter.

\begin{figure}[t]
\centering
\epsscale{1.0}
\plotone{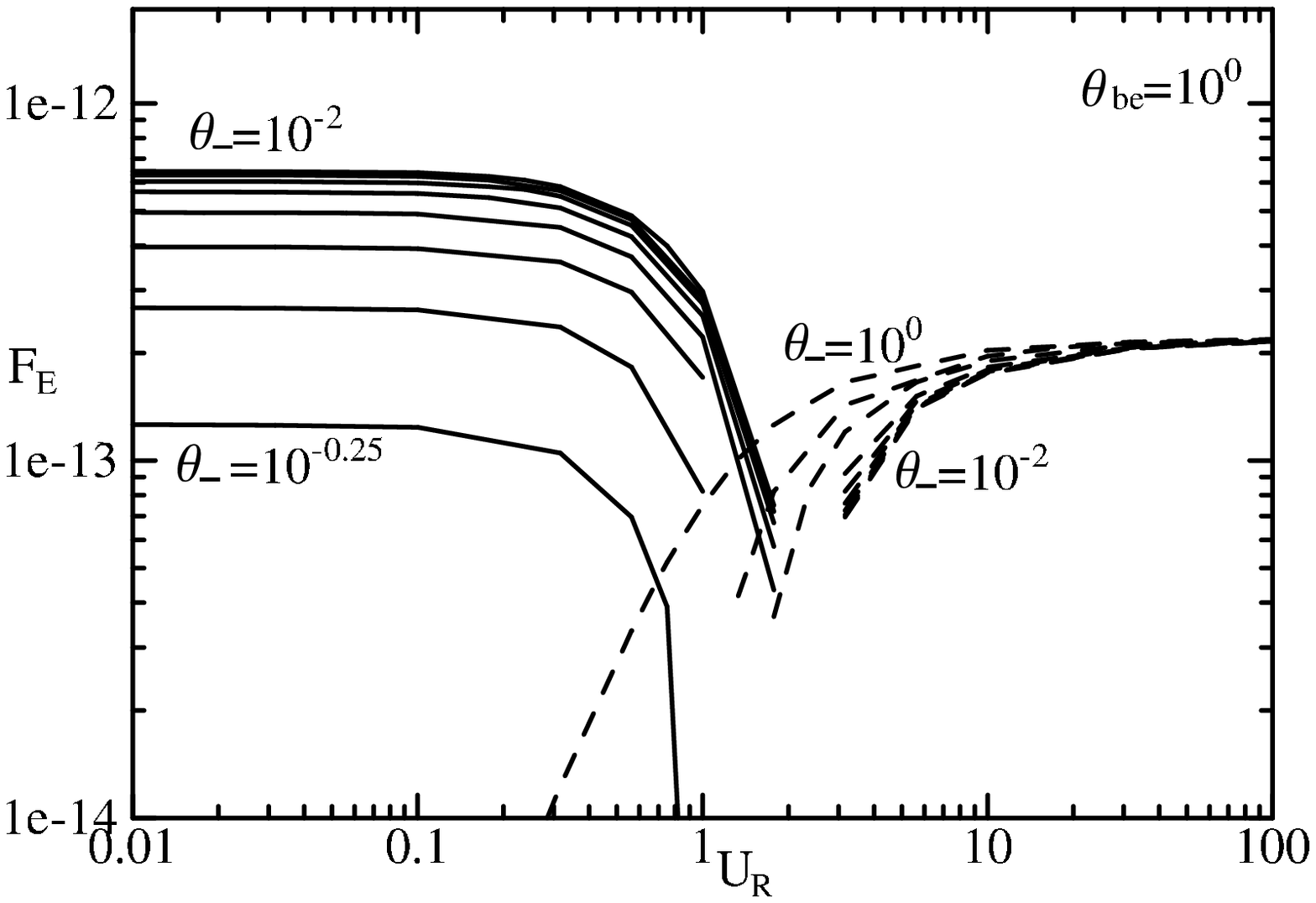}
\caption{
$F_{\rm E}$ of e${^-}$-e${^-}$
interaction for $\theta_{\rm be}=1$ in cgs unit. 
The temperatures of the outflowing
electrons, $\theta_{\rm -}$,
are from $10^{-2}$ to $10^0$ at $10^{0.25}$
intervals in logarithmic scale.
Dashed lines are $|F_{\rm E}|$ for negative $F_{\rm E}$.
}
\end{figure}

\begin{figure}[t]
\centering
\epsscale{1.0}
\plotone{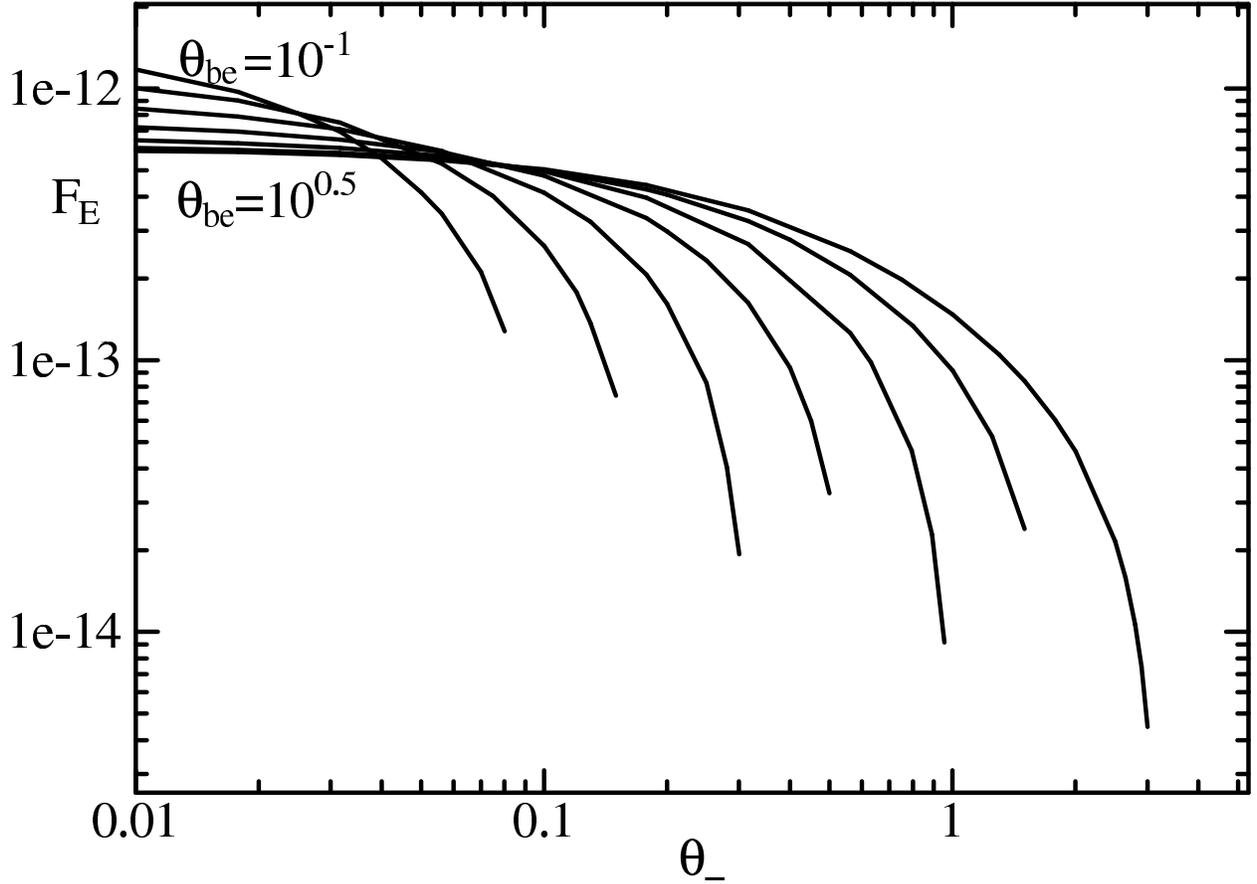}
\caption{
$F_{\rm E}$ of e${^-}$-e${^-}$
interaction for $U_{\rm R}=0$ in cgs unit. 
The temperatures of the background
electrons, $\theta_{\rm be}$,
are from $10^{-1}$ to $10^{0.5}$ at $10^{0.25}$
intervals in logarithmic scale.
}
\end{figure}

Figure 7 shows ET between two electron fluids
for $\theta_{\rm be}=1$.
For $\theta_- \lesssim 0.1$ the functional shapes of ET are almost invariant,
because we can neglect the bulk motion for $U_{\rm R} \ll \theta_{\rm be}$.
From the figure we can see ET becomes zero at $U_{\rm R} \sim $ a few,
although it is difficult to
determine the value of $U_{\rm R}$, at which ET becomes zero,
within our computational precision.
As $\theta_-$ increases, $|F_{\rm E}|$ also increases in cases of negative ET
($U_{\rm R} \gtrsim 1$),
while $|F_{\rm E}|$ decreases for positive ET ($U_{\rm R} \lesssim 1$).

When $U_{\rm R} \gg \theta_{\rm be}$, the thermal motion can be neglected.
In this case, from equation (\ref{de}),
the energy loss of outflowing electrons in one collision
$\langle \Delta E \rangle \sim U_{\rm R} m_{\rm e} c^2 \sin^2{(\alpha/2)}$.
Since the mass of the colliding particle is the same,
the Lorentz factor of electrons
in the CM frame $\gamma \sim \gamma_{\rm CM} \propto U_{\rm R}^{1/2}$.
The effective cross section $\sigma_{\rm eff} \propto \gamma^{-2}
\propto U_{\rm R}^{-1}$ so that $F_{\rm E}$ becomes constant
for $U_{\rm R} \gg \theta_{\rm be}$ as shown in Figure 7.

We fit our results by functions,
\begin{eqnarray}
F_{\rm E}=A_{\rm L} \exp{\left[-\left( U_{\rm R}/U_{\rm L} \right)^2 \right]},
\label{feee1}
\end{eqnarray}
for $U_{\rm R} \lesssim 1$ and
\begin{eqnarray}
F_{\rm E}=-A_{\rm H} \exp{\left[-\left(U_{\rm R}/U_{\rm H} \right)^p \right]},
\label{feee2}
\end{eqnarray}
for $U_{\rm R} \gtrsim 1$.
Fitting results are listed in Tables 11 and 12.

We tabulate the fitting parameters only for
$\theta_{\rm be}=0.1$ and $1$.
However, as is shown in Figure 7, $F_{\rm E}$ is almost constant
for $U_{\rm R} \ll 1$.
Thus, it is valuable to estimate $F_{\rm E}$ for $U_{\rm R}=0$
with a wide range of $\theta_{\rm be}$.
Figure 8 shows the results, which are fitted by a function
\begin{eqnarray}
F_{\rm E}=A \exp{\left[-\left(\theta_-/\theta_0 \right)^p \right]},
\label{feee3}
\end{eqnarray}
(see Table 13).
Although the $\theta_{\rm be}$-dependece is hardly expressed by a simple formula,
we can interpolate $F_{\rm E}$ from Table 13.

\subsection{MOMENTUM TRANSFER in e-e INTERACTION}

\begin{figure}[t]
\centering
\epsscale{1.0}
\plotone{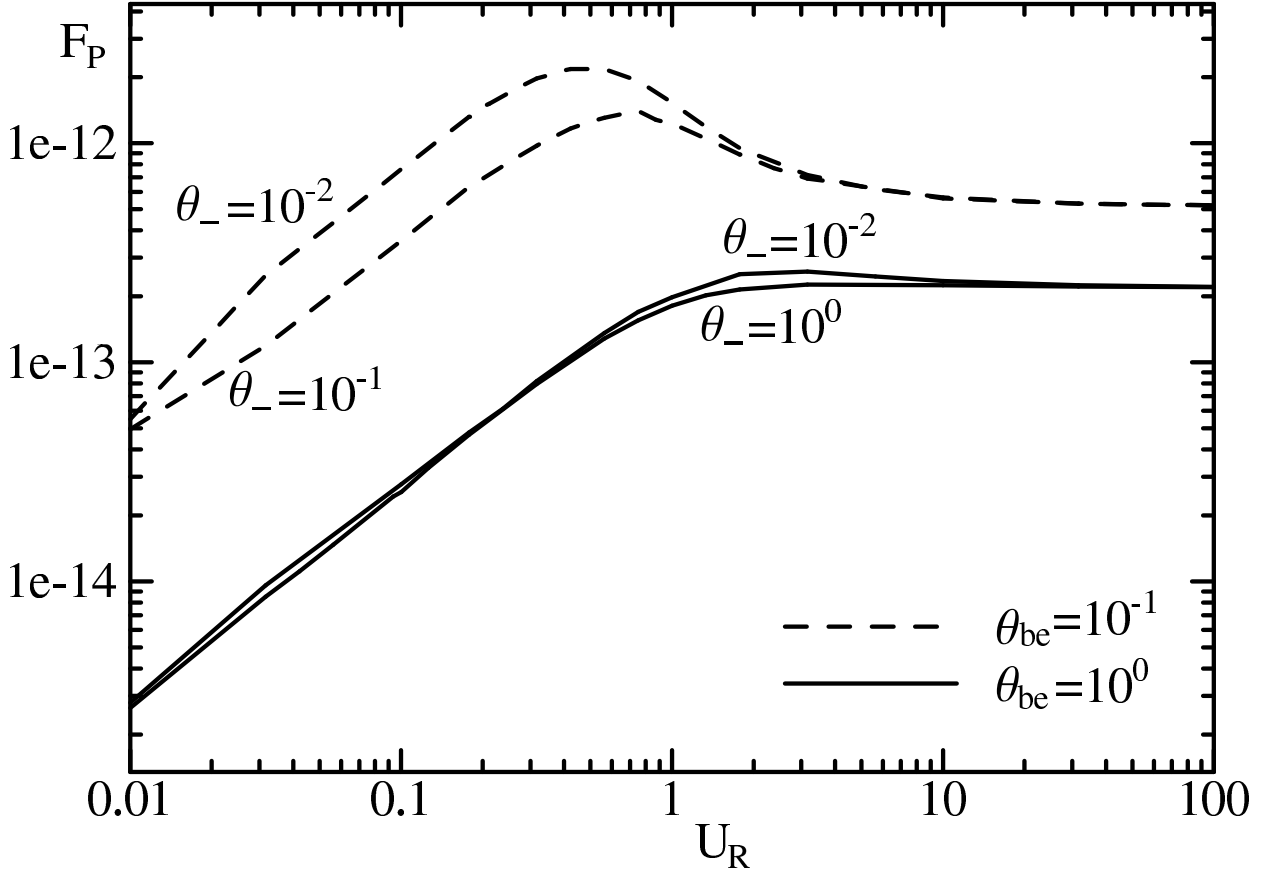}
\caption{
$F_{\rm P}$ of e${^-}$-e${^-}$
interaction in cgs unit. 
}
\end{figure}

\begin{figure}[t]
\centering
\epsscale{1.0}
\plotone{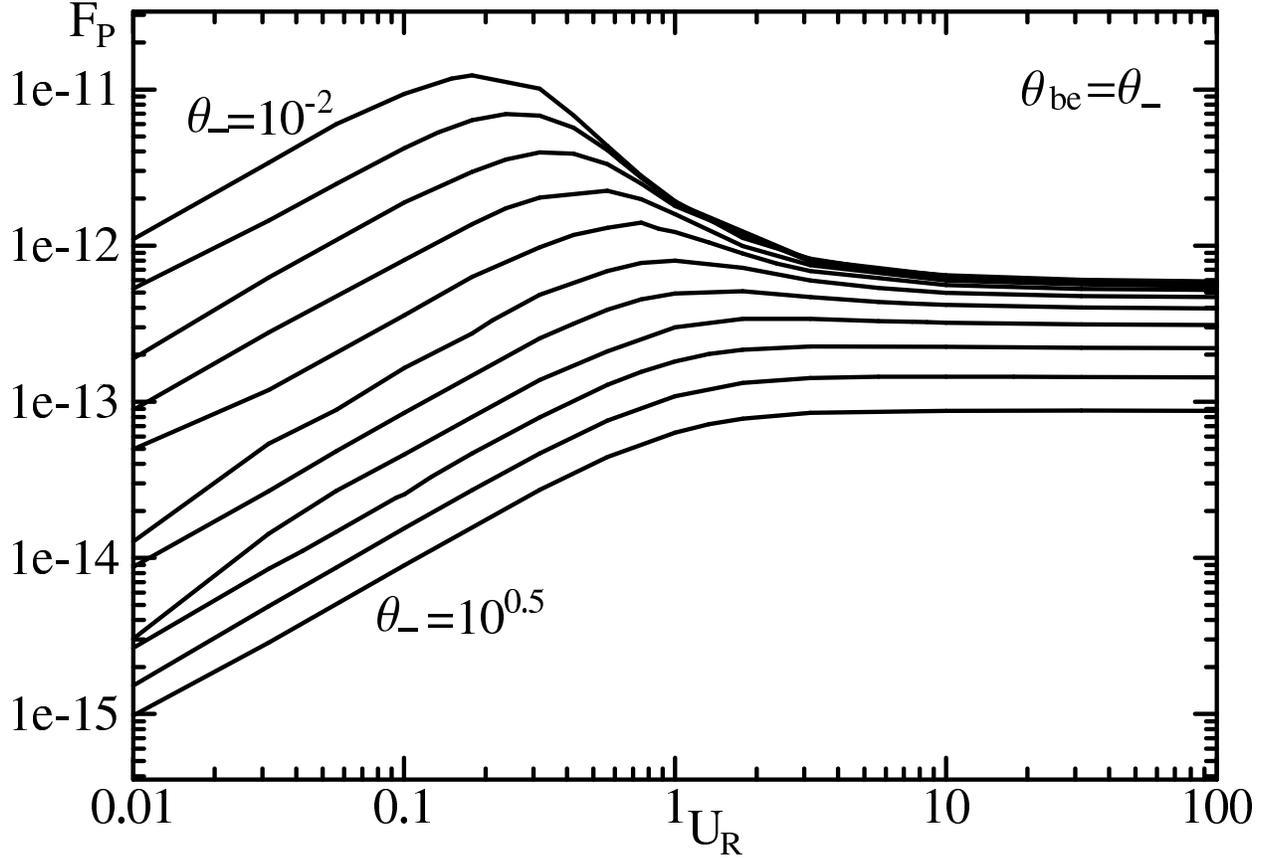}
\caption{
$F_{\rm P}$ of e${^-}$-e${^-}$
interaction for $\theta_{\rm be}=\theta_{\rm -}$ in cgs unit.
The temperatures
are from $10^{-2}$ to $10^{0.5}$ at $10^{0.25}$
intervals in logarithmic scale.
}
\end{figure}

As is the case with energy transfer,
the momentum transfers for e${^-}$-e${^-}$ and e${^-}$-e${^+}$ interactions
are almost the same.
The numerical results of $F_{\rm P}$ are plotted in Figure 9.
For $\theta_{\rm be}=10^0$ the results are almost
independent of the temperature of the outflowing electrons $\theta_-$.
Even for $\theta_{\rm be}=10^{-1}$,
$F_{\rm P}$ decreases only by a factor of less than 2 with
increasing $\theta_-=10^{-2}$ to $10^{-1}$.
Figure 9 clearly shows the reduction of $F_{\rm P}$ due to increasing $\theta_{\rm be}$.
As we have seen in the former cases,
the growth of the average $\gamma$ by rising $\theta_{\rm be}$
results in the reduction of the cross section ($\sigma_{\rm eff}
\propto \gamma^{-2} \beta^{-4}$ for relativistic case).
For $U_{\rm R} \gg 1$, $F_{\rm P}$ is almost constant
differently from the case of p-e interaction.
This is because $\beta_{\rm CM} \sim 1$ in e-e interaction
and a resultant relation $\langle \Delta p \rangle \sim \langle \Delta E \rangle/c$.
Therefore, our results show $F_{\rm P} \simeq -F_{\rm E} \propto U_{\rm R}^0$
for $U_{\rm R} \gg \theta_{\rm be}$.

In contrast to p-e interaction,
the cooling and heating properties of the two fluids in this case
are the same.
Therefore, if the flow velocity is non-relativistic,
the two temperatures may be the same.
Thus we calculate $F_{\rm P}$ mainly with $\theta_-=\theta_{\rm be}$,
as shown in Figure 10.
Even if the two temperatures are different,
the correction is not so large as shown in Figure 9.
We fit our results by functions;

\begin{eqnarray}
F_{\rm P}=\left\{
\begin{array}{ll}
A_{\rm L} \exp{\left[-\left( \frac{U_{\rm R}}{U_{\rm L}} \right)^p \right]}
\left( \frac{U_{\rm R}}{0.01} \right)^q,
& \mbox {for $U_{\rm R} < U_{\rm th}$},\\
A_{\rm H} \exp{\left[-\left( \frac{U_{\rm R}}{U_{\rm H}} \right)^r \right]},
& \mbox {for $U_{\rm R} \geq U_{\rm th}$},
\end{array} \right.
\end{eqnarray}
and results are summarized in Table 14.
The symbol $\infty$ for $U_{\rm H}$ in this table
means that $F_{\rm P}$ is constant as $F_{\rm P}=A_{\rm H}$
for $U_{\rm R} \geq U_{\rm th}$.

\subsection{MOMENTUM TRANSFER in PAIR-ANNIHILATION}

We shortly comment on the effect of pair-annihilation process here.
The outflowing positrons may interact with the background electrons
and turn into gamma-rays.
This process decreases the momentum of the outflowing fluid,
and it is not trivial whether we can neglect this momentum loss
process or not.
In Figure 11 we plot $F_{\rm P}$ of e${^-}$-e${^+}$
due to pair-annihilation.
As $\theta_-$ increases for $\theta_{\rm be}=10^0$,
the momentum loss due to pair-annihilation increases,
while that due to pair-scattering is almost constant.
Even for $\theta_-=10^0$, the momentum loss due to
pair-annihilation is smaller than that due
to pair-scattering by one order of magnitude.

For smaller values of $\theta_{\rm be}<10^0$,
the contribution of pair-annihilation to $F_{\rm P}$
becomes much more negligible.
Therefore, we can neglect the effect of pair-annihilation
on the flow within precision we require.

\begin{figure}[t]
\centering
\epsscale{1.0}
\plotone{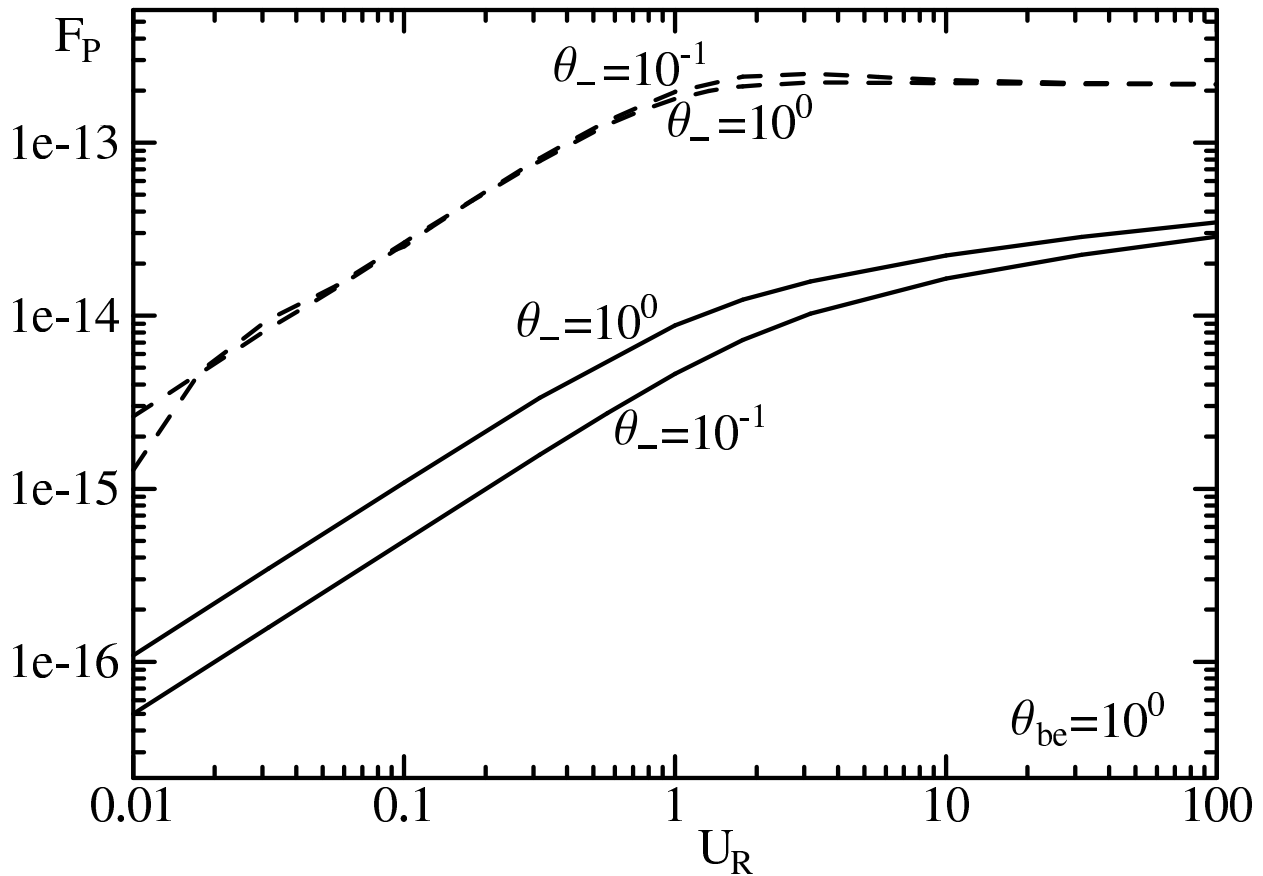}
\caption{
$F_{\rm P}$ of e${^-}$-e${^+}$ interaction
due to scattering (dashed) and pair-annihilation (solid)
for $\theta_{\rm be}=10^0$ in cgs unit.
}
\end{figure}

\section{SUMMARY}

Motivated by electron-positron outflows from AGNs,
we numerically calculate the energy and momentum transfer rates
due to Coulomb scattering
between two fluids with a relative velocity
for plausible parameters in AGN models.
Although several plasma effects, such as two-stream instability etc.,
may enhance the energy and momentum transfers,
the effects of Coulomb scattering are the first to be taken
into account.
Our tables obtained from the numerical results
are useful to simulate pair outflows from hot plasmas,
or evaluate the interaction between AGN jets and the ambient medium.
Especially, the momentum transfer rate is indispensable
for such simulations.
Using the results in this paper,
a simulation in \citet{asa06} shows that the frictional force due to
Coulomb scattering is comparable to radiative force
for plausible parameter sets in AGN jet models.
Therefore, we cannot neglect the effects of Coulomb scattering
in such simulations.

\begin{acknowledgments}
We thank the anonymous referee for his useful comments.
This work is partially supported by Scientific Research Grants
(F.T. 14079205 and 16540215) from the Ministry of Education, Culture,
Science and Technology of Japan.
\end{acknowledgments}

\onecolumn

\begin{table}[h]
\begin{center}
\caption{Fitting parameters of $F_{\rm E}$ for p$-$e interaction, where $\Theta_{\rm p}=
10^{0.75}$. See eq. (\ref{fepe}).}
\begin{tabular}{lcccccc}
\hline \hline
$\theta_{\rm e}$ & $A_{\rm L}$ & $U_0$ & $p$ &
$A_{\rm H}$ & $r$ & $U_{\rm th}$ \\ \hline
$10^{-2}$ & 9.6e-13 & 0.166 & 2.0 & 3.0e-15 & -1.2 & 0.35 \\ 
$10^{-1.75}$ & 4.8e-13 & 0.209 & 2.0 & 3.1e-15 & -1.2 & 0.42 \\ 
$10^{-1.5}$ & 2.3e-13 & 0.275 & 2.0 & 3.1e-15 & -1.2 & 0.51 \\ 
$10^{-1.25}$ & 1.0e-13 & 0.369 & 2.0 & 3.1e-15 & -1.3 & 0.63 \\ 
$10^{-1}$ & 4.4e-14 & 0.508 & 2.0 & 3.5e-15 & -1.3 & 0.75 \\ 
$10^{-0.75}$ & 2.0e-14 & 0.673 & 2.0 & 4.2e-15 & -1.5 & 0.59 \\ 
$10^{-0.5}$ & 8.3e-15 & 1.008 & 1.5 & - & - & - \\ 
$10^{-0.25}$ & 3.6e-15 & 1.443 & 1.3 & - & - & - \\ 
$10^{0.0}$ & 1.7e-15 & 1.511 & 1.3 & - & - & - \\ \hline
\end{tabular}
\end{center}
\end{table}

\begin{table}[h]
\begin{center}
\caption{Same as Table 1 but for $\Theta_{\rm p}=
10$}
\begin{tabular}{lccccc}
\hline \hline
$\theta_{\rm e}$ & $A_{\rm L}$ & $U_0$ & $A_{\rm H}$ & $r$ & $U_{\rm th}$ \\ \hline
$10^{-2}$ & 1.4e-12 & 0.179 & 5.6e-15 & -1.2 & 0.37 \\ 
$10^{-1.75}$ & 7.3e-13 & 0.224 & 5.6e-15 & -1.2 & 0.44 \\ 
$10^{-1.5}$ & 3.7e-13 & 0.284 & 5.8e-15 & -1.2 & 0.52 \\ 
$10^{-1.25}$ & 1.7e-13 & 0.374 & 5.9e-15 & -1.2 & 0.63 \\ 
$10^{-1}$ & 7.7e-14 & 0.515 & 6.2e-15 & -1.2 & 0.76 \\ 
$10^{-0.75}$ & 3.5e-14 & 0.698 & 7.7e-15 & -1.4 & 0.69 \\ 
$10^{-0.5}$ & 1.5e-14 & 0.909 & 5.2e-15 & -1.4 & 0.76 \\ 
$10^{-0.25}$ & 7.0e-15 & 1.161 & 3.6e-15 & -1.4 & 1.01 \\ 
$10^{0}$ & 3.2e-15 & 1.544 & 2.3e-15 & -1.4 & 1.31 \\ 
$10^{0.25}$ & 1.6e-15 & 1.359 & 9.8e-16 & -1.4 & 1.16 \\ 
$10^{0.5}$ & 7.1e-16 & 1.209 & - & - & - \\ \hline
\end{tabular}
\end{center}
\end{table}

\begin{table}[h]
\begin{center}
\caption{Same as Table 1 but for $\Theta_{\rm p}=
10^{1.5}$}
\begin{tabular}{lccccc}
\hline \hline
$\theta_{\rm e}$ & $A_{\rm L}$ & $U_0$ & $A_{\rm H}$ & $r$ & $U_{\rm th}$ \\ \hline
$10^{-2}$ & 1.8e-12 & 0.239 & 1.9e-14 & -1.2 & 0.45 \\ 
$10^{-1.75}$ & 1.3e-12 & 0.274 & 2.0e-14 & -1.2 & 0.50 \\ 
$10^{-1.5}$ & 7.7e-13 & 0.303 & 2.0e-14 & -1.2 & 0.51 \\ 
$10^{-1.25}$ & 4.1e-13 & 0.425 & 2.3e-14 & -1.2 & 0.66 \\ 
$10^{-1}$ & 2.1e-13 & 0.545 & 2.2e-14 & -1.2 & 0.75 \\ 
$10^{-0.75}$ & 9.9e-14 & 0.778 & 2.4e-14 & -1.3 & 0.87 \\ 
$10^{-0.5}$ & 4.7e-14 & 0.952 & 1.7e-14 & -1.3 & 0.79 \\ 
$10^{-0.25}$ & 2.2e-14 & 1.230 & 1.3e-14 & -1.3 & 1.04 \\ 
$10^{0.0}$ & 1.1e-14 & 1.479 & 6.9e-15 & -1.2 & 0.96 \\ 
$10^{0.25}$ & 5.6e-15 & 1.766 & 4.6e-15 & -1.3 & 1.19 \\ 
$10^{0.5}$ & 2.9e-15 & 1.783 & 2.5e-15 & -1.3 & 1.42 \\ \hline
\end{tabular}
\end{center}
\end{table}

\begin{table}[h]
\begin{center}
\caption{Same as Table 1 but for $\Theta_{\rm p}=
10^{2}$}
\begin{tabular}{lccccc}
\hline \hline
$\theta_{\rm e}$ & $A_{\rm L}$ & $U_0$ & $A_{\rm H}$ & $r$ & $U_{\rm th}$ \\ \hline
$10^{-2}$ & 1.6e-12 & 0.383 & 6.1e-14 & -1 & 0.65 \\ 
$10^{-1.75}$ & 1.4e-12 & 0.408 & 6.2e-14 & -1.1 & 0.66 \\ 
$10^{-1.5}$ & 1.1e-12 & 0.446 & 6.4e-14 & -1.1 & 0.69 \\ 
$10^{-1.25}$ & 7.3e-13 & 0.524 & 6.9e-14 & -1.2 & 0.74 \\ 
$10^{-1}$ & 4.4e-13 & 0.644 & 7.3e-14 & -1.2 & 0.80 \\ 
$10^{-0.75}$ & 2.5e-13 & 0.825 & 6.0e-14 & -1.2 & 0.96 \\ 
$10^{-0.5}$ & 1.3e-13 & 0.933 & 5.1e-14 & -1.2 & 0.75 \\ 
$10^{-0.25}$ & 6.4e-14 & 1.245 & 3.5e-14 & -1.1 & 0.95 \\ 
$10^{0}$ & 3.3e-14 & 1.441 & 2.2e-14 & -1.1 & 1.10 \\ 
$10^{0.25}$ & 1.7e-14 & 1.630 & 1.3e-14 & -1.1 & 1.23 \\ 
$10^{0.5}$ & 9.4e-15 & 1.816 & 7.8e-15 & -1.1 & 1.37 \\ \hline
\end{tabular}
\end{center}
\end{table}

\begin{table}[h]
\begin{center}
\caption{Same as Table 1 but for $\Theta_{\rm p}=
10^{2.5}$}
\begin{tabular}{lcccccc}
\hline \hline
$\theta_{\rm e}$ & $A_{\rm L}$ & $U_0$ & $p$ &
$A_{\rm H}$ & $r$ & $U_{\rm th}$ \\ \hline
$10^{-2}$ & 1.1e-12 & 0.673 & 2.0 & 2.2e-13 & -1.1 & 0.79 \\ 
$10^{-1.75}$ & 1.0e-12 & 0.704 & 2.0  & 2.1e-13 & -1.1 & 0.83 \\ 
$10^{-1.5}$ & 9.4e-13 & 0.732 & 2.0  & 2.1e-13 & -1.1 & 0.83 \\ 
$10^{-1.25}$ & 8.0e-13 & 0.782 & 2.0  & 2.1e-13 & -1.1 & 0.84 \\ 
$10^{-1}$ & 6.2e-13 & 0.867 & 2.0  & 2.0e-13 & -1.1 & 0.86 \\ 
$10^{-0.75}$ & 4.3e-13 & 0.992 & 2.0  & 1.7e-13 & -1.1 & 0.87 \\ 
$10^{-0.5}$ & 2.7e-13 & 1.164 & 2.0  & 1.3e-13 & -1.1 & 1.05 \\ 
$10^{-0.25}$ & 1.5e-13 & 1.540 & 1.6  & 9.5e-14 & -1.1 & 1.36 \\ 
$10^{0}$ & 8.5e-14 & 1.757 & 1.6  & 6.5e-14 & -1.1 & 1.45 \\ 
$10^{0.25}$ & 4.7e-14 & 1.924 & 1.6  & 4.1e-14 & -1.1 & 1.73 \\ 
$10^{0.5}$ & 2.6e-14 & 1.994 &  1.6 & 2.4e-14 & -1.1 & 1.68 \\ \hline
\end{tabular}
\end{center}
\end{table}

\begin{table}[h]
\begin{center}
\caption{Fitting parameters of $F_{\rm P}$ for p$-$e interaction, where $\Theta_{\rm p}=
10^{0.75}$. See eq. (\ref{fppe}).}
\begin{tabular}{lccccccc}
\hline \hline
$\theta_{\rm e}$ & $A_{\rm L}$ & $U_0$ & $p$ & $q$ &
$A_{\rm H}$ & $r$ & $U_{\rm th}$ \\ \hline
$10^{-2}$ & 1.1e-12 & 0.214 & 1.5 & 1.0 & 8.7e-13 & -1.1 & 0.48 \\ 
$10^{-1.75}$ & 4.6e-13 & 0.255 & 1.5 & 1.1 & 7.8e-13 & -1.1 & 0.56 \\ 
$10^{-1.5}$ & 2.2e-13 & 0.322 & 1.5 & 1.1 & 7.8e-13 & -1.1 & 0.65 \\ 
$10^{-1.25}$ & 1.1e-13 & 0.502 & 1.5 & 1.0 & 7.8e-13 & -1.1 & 0.90 \\ 
$10^{-1}$ & 5.0e-14 & 0.631 & 1.5 & 1.0 & 5.6e-13 & -1.0 & 1.16 \\ 
$10^{-0.75}$ & 1.7e-14 & 0.700 & 1.3 & 1.1 & 5.6e-13 & -1.0 & 1.02 \\ 
$10^{-0.5}$ & 5.9e-15 & 0.713 & 1.0 & 1.2 & 4.1e-13 & -1.0 & 2.02 \\  
$10^{-0.25}$ & 4.0e-15 & 0.572 & 0.8 & 1.2 & 3.1e-13 & -1.0 & 2.29 \\  
$10^{0.0}$ & 2.3e-15 & 0.798 & 0.8 & 1.1 & 2.2e-13 & -1.0 & 2.70 \\  \hline
\end{tabular}
\end{center}
\end{table}

\begin{table}[h]
\begin{center}
\caption{Same as Table 6 but for $\Theta_{\rm p}=
10$}
\begin{tabular}{lccccccc}
\hline \hline
$\theta_{\rm e}$ & $A_{\rm L}$ & $U_0$ & $p$ & $q$ &
$A_{\rm H}$ & $r$ & $U_{\rm th}$ \\ \hline
$10^{-2}$ & 8.6e-13 & 0.237 & 1.5 & 1.0 & 1.1e-12 & -1.2 & 0.44 \\ 
$10^{-1.75}$ & 4.0e-13 & 0.257 & 1.5 & 1.1 & 8.1e-13 & -1.1 & 0.53 \\ 
$10^{-1.5}$ & 1.9e-13 & 0.341 & 1.5 & 1.1 & 7.9e-13 & -1.1 & 0.66 \\ 
$10^{-1.25}$ & 1.1e-13 & 0.488 & 1.5 & 1.0 & 7.5e-13 & -1.1 & 0.85 \\ 
$10^{-1}$ & 3.1e-14 & 0.512 & 1.5 & 1.2 & 6.8e-13 & -1.1 & 0.74 \\ 
$10^{-0.75}$ & 1.8e-14 & 0.914 & 1.5 & 1.0 & 5.6e-13 & -1.0 & 0.88 \\ 
$10^{-0.5}$ & 9.5e-15 & 1.074 & 1.5 & 1.0 & 4.1e-13 & -1.0 & 1.52 \\  
$10^{-0.25}$ & 4.7e-15 & 1.195 & 1.2 & 1.0 & 3.0e-13 & -1.0 & 1.86 \\  
$10^{0.0}$ & 1.8e-15 & 1.090 & 1.0 & 1.1 & 1.7e-13 & -0.9 & 2.75 \\  
$10^{0.25}$ & 1.2e-15 & 1.418 & 1.0 & 1.0 & 1.2e-13 & -0.9 & 3.23 \\  
$10^{0.5}$ & 9.9e-16 & 1.617 & 1.0 & 0.9 & 6.9e-14 & -0.9 & 3.44 \\  \hline
\end{tabular}
\end{center}
\end{table}

\begin{table}[h]
\begin{center}
\caption{Same as Table 6 but for $\Theta_{\rm p}=
10^{1.5}$}
\begin{tabular}{lccccccc}
\hline \hline
$\theta_{\rm e}$ & $A_{\rm L}$ & $U_0$ & $p$ & $q$ &
$A_{\rm H}$ & $r$ & $U_{\rm th}$ \\ \hline
$10^{-2}$ & 3.6e-13 & 0.329 & 1.5 & 1.0 & 7.7e-13 & -1.1 & 0.69 \\ 
$10^{-1.75}$ & 2.5e-13 & 0.359 & 1.5 & 1.0 & 7.6e-13 & -1.1 & 0.70 \\ 
$10^{-1.5}$ & 1.4e-13 & 0.359 & 1.5 & 1.1 & 7.5e-13 & -1.1 & 0.64 \\ 
$10^{-1.25}$ & 8.1e-14 & 0.538 & 1.5 & 1.0 & 7.2e-13 & -1.1 & 0.92 \\ 
$10^{-1}$ & 4.0e-14 & 0.676 & 1.5 & 1.0 & 6.6e-13 & -1.1 & 1.00 \\ 
$10^{-0.75}$ & 1.9e-14 & 0.866 & 1.5 & 1.0 & 5.3e-13 & -1.0 & 1.15 \\ 
$10^{-0.5}$ & 1.1e-14 & 1.271 & 1.8 & 0.9 & 4.0e-13 & -1.0 & 1.32 \\  
$10^{-0.25}$ & 4.6e-15 & 1.187 & 1.2 & 1.0 & 2.9e-13 & -1.0 & 1.82 \\  
$10^{0.0}$ & 2.0e-15 & 1.600 & 1.2 & 1.0 & 2.1e-13 & -1.0 & 3.24 \\  
$10^{0.25}$ & 1.2e-15 & 1.560 & 1.2 & 1.0 & 1.1e-13 & -0.9 & 2.29 \\  
$10^{0.5}$ & 5.5e-16 & 1.191 & 1.0 & 1.1 & 6.8e-14 & -0.9 & 2.39 \\  \hline
\end{tabular}
\end{center}
\end{table}

\begin{table}[h]
\begin{center}
\caption{Same as Table 6 but for $\Theta_{\rm p}=
10^{2}$}
\begin{tabular}{lccccccc}
\hline \hline
$\theta_{\rm e}$ & $A_{\rm L}$ & $U_0$ & $p$ & $q$ &
$A_{\rm H}$ & $r$ & $U_{\rm th}$ \\ \hline
$10^{-2}$ & 1.0e-13 & 0.478 & 1.5 & 1.0 & 7.2e-13 & -1.1 & 0.81 \\ 
$10^{-1.75}$ & 7.7e-14 & 0.527 & 1.5 & 1.0 & 7.1e-13 & -1.1 & 0.84 \\ 
$10^{-1.5}$ & 5.4e-14 & 0.463 & 1.5 & 1.1 & 6.7e-13 & -1.1 & 0.61 \\ 
$10^{-1.25}$ & 5.7e-14 & 0.697 & 1.5 & 0.9 & 5.3e-13 & -1.0 & 0.75 \\ 
$10^{-1}$ & 2.0e-14 & 0.677 & 1.5 & 1.1 & 5.7e-13 & -1.0 & 0.85 \\ 
$10^{-0.75}$ & 1.5e-14 & 0.876 & 1.3 & 1.0 & 4.9e-13 & -1.0 & 1.23 \\ 
$10^{-0.5}$ & 8.0e-15 & 1.047 & 1.3 & 1.0 & 3.6e-13 & -1.0 & 1.56 \\  
$10^{-0.25}$ & 4.0e-15 & 1.229 & 1.3 & 1.0 & 2.6e-13 & -1.0 & 1.74 \\  
$10^{0.0}$ & 2.3e-15 & 1.255 & 1.0 & 1.0 & 1.9e-13 & -1.0 & 2.51 \\  
$10^{0.25}$ & 8.7e-16 & 1.139 & 1.0 & 1.1 & 9.5e-14 & -0.9 & 2.59 \\  
$10^{0.5}$ & 6.7e-16 & 1.358 & 1.0 & 1.0 & 6.3e-14 & -0.9 & 2.63 \\  \hline
\end{tabular}
\end{center}
\end{table}

\begin{table}[h]
\begin{center}
\caption{Same as Table 6 but for $\Theta_{\rm p}=
10^{2.5}$}
\begin{tabular}{lccccccc}
\hline \hline
$\theta_{\rm e}$ & $A_{\rm L}$ & $U_0$ & $p$ & $q$ &
$A_{\rm H}$ & $r$ & $U_{\rm th}$ \\ \hline
$10^{-2}$ & 1.8e-14 & 0.783 & 1.5 & 1.0 & 4.6e-13 & -1.0 & 0.97 \\ 
$10^{-1.75}$ & 1.9e-14 & 1.008 & 2.0 & 0.9 & 4.5e-13 & -1.0 & 0.99 \\ 
$10^{-1.5}$ & 1.3e-14 & 0.925 & 2.0 & 1.0 & 4.3e-13 & -1.0 & 0.94 \\ 
$10^{-1.25}$ & 1.4e-14 & 1.102 & 2.5 & 0.9 & 4.0e-13 & -1.0 & 0.99 \\ 
$10^{-1}$ & 9.1e-15 & 1.042 & 1.5 & 1.0 & 4.0e-13 & -1.0 & 1.29 \\ 
$10^{-0.75}$ & 6.5e-15 & 1.178 & 1.5 & 1.0 & 3.6e-13 & -1.0 & 1.45 \\ 
$10^{-0.5}$ & 4.5e-15 & 1.218 & 1.5 & 1.0 & 2.8e-13 & -1.0 & 1.51 \\  
$10^{-0.25}$ & 2.8e-15 & 1.298 & 1.1 & 1.0 & 2.2e-13 & -1.0 & 2.57 \\  
$10^{0.0}$ & 1.6e-15 & 1.436 & 1.2 & 1.0 & 1.3e-13 & -0.9 & 2.11 \\  
$10^{0.25}$ & 7.3e-16 & 1.124 & 1.0 & 1.1 & 8.5e-14 & -0.9 & 2.32 \\  
$10^{0.5}$ & 4.7e-16 & 0.916 & 0.8 & 1.1 & 5.1e-14 & -0.9 & 2.89 \\  \hline
\end{tabular}
\end{center}
\end{table}

\begin{table}[h]
\begin{center}
\caption{Fitting parameters of $F_{\rm E}$ for e$-$e interaction,
where $\theta_{\rm be}=10^{-1}$. See eqs. (\ref{feee1}) and (\ref{feee2}).}
\begin{tabular}{lccccc}
\hline \hline
$\theta_{\rm -}$ & $A_{\rm L}$ & $U_{\rm L}$ & $A_{\rm H}$ &
$U_{\rm H}$ & $p$  \\ \hline
$10^{-2}$ & 1.2e-12 & 0.315 & 5.2e-13 & 0.5 & -3 \\ 
$10^{-1.75}$ & 9.7e-13 & 0.313 & 5.2e-13 & 0.5 & -3 \\ 
$10^{-1.5}$ & 6.8e-13 & 0.321 & 5.2e-13 & 0.5 & -3 \\ 
$10^{-1.25}$ & 3.4e-13 & 0.265 & 5.2e-13 & 0.5 & -3 \\ \hline
\end{tabular}
\end{center}
\end{table}

\begin{table}[h]
\begin{center}
\caption{Same as Table 11 but for $\theta_{\rm be}=
10^0$}
\begin{tabular}{lccccc}
\hline \hline
$\theta_{\rm -}$ & $A_{\rm L}$ & $U_{\rm L}$ & $A_{\rm H}$ &
$U_{\rm H}$ & $p$  \\ \hline
$10^{-2}$ & 6.5e-13 & 1.117 & 2.2e-13 & 3.469 & -2 \\ 
$10^{-1.75}$ & 6.3e-13 & 1.124 & 2.2e-13 & 3.444 & -2 \\ 
$10^{-1.5}$ & 6.1e-13 & 1.123 & 2.2e-13 & 3.490 & -2 \\ 
$10^{-1.25}$ & 5.7e-13 & 1.112 & 2.2e-13 & 3.218 & -1.5 \\ 
$10^{-1}$ & 5.0e-13 & 1.104 & 2.2e-13 & 2.866 & -1.5 \\ 
$10^{-0.75}$ & 4.0e-13 & 1.076 & 2.2e-13 & 2.881 & -1.5 \\ 
$10^{-0.5}$ & 2.7e-13 & 0.915 & 2.2e-13 & 2.408 & -1.5 \\ 
$10^{-0.25}$ & 1.3e-13 & 0.679 & 2.2e-13 & 1.805 & -1.5 \\ \hline
\end{tabular}
\end{center}
\end{table}

\begin{table}[h]
\begin{center}
\caption{Fitting parameters of $F_{\rm E}$ for e$-$e interaction
for $U_{\rm R}=0$. See eq. (\ref{feee3}).}
\begin{tabular}{lccc}
\hline \hline
$\theta_{\rm be}$ & $A$ & $\theta_0$  & $p$  \\ \hline
$10^{-1}$ & 1.2e-12 & 0.049 & 1.5 \\ 
$10^{-0.75}$ & 1.0e-12 & 0.080 & 1.4 \\ 
$10^{-0.5}$ & 8.5e-13 & 0.13 & 1.3 \\ 
$10^{-0.25}$ & 7.2e-13 & 0.22 & 1.3 \\
$10^{0.0}$ & 6.5e-13 & 0.35 & 1.2 \\  
$10^{0.25}$ & 6.1e-13 & 0.52 & 1.0 \\  
$10^{0.5}$ & 5.9e-13 & 0.7 & 1.0 \\  \hline
\end{tabular}
\end{center}
\end{table}

\begin{table}[h]
\begin{center}
\caption{Fitting parameters of $F_{\rm P}$ for e$-$e interaction
for $\theta_-=\theta_{\rm be}$. See eq. (\ref{feee3}).}
\begin{tabular}{lcccccccc}
\hline \hline
$\theta_{\rm e}$ & $A_{\rm L}$ & $U_{\rm L}$ & $p$ & $q$ &
$A_{\rm H}$ & $U_{\rm H}$ & $r$ & $U_{\rm th}$ \\ \hline
$10^{-2}$ & 1.1e-12 & 0.273 & 2.0 & 1.0 & 5.9e-13 & 1.111 & -0.85 & 0.36 \\ 
$10^{-1.75}$ & 5.3e-13 & 0.428 & 2.0 & 0.9 & 5.9e-13 & 1.093 & -0.9 & 0.55 \\ 
$10^{-1.5}$ & 1.9e-13 & 0.511 & 2.0 & 1.0 & 5.7e-13 & 1.023 & -1.0 & 1.16 \\ 
$10^{-1.25}$ & 8.9e-14 & 0.676 & 1.5 & 1.0 & 5.5e-13 & 1.059 & -1.0 & 0.89 \\ 
$10^{-1}$ & 5.0e-14 & 1.081 & 1.5 & 0.9 & 5.2e-13 & 0.888 & -1.0 & 1.38 \\ 
$10^{-0.75}$ & 1.3e-14 & 0.650 & 0.75 & 1.2 & 4.7e-13 & 0.762 & -1.0 & 2.33 \\ 
$10^{-0.5}$ & 8.8e-15 & 0.931 & 0.7 & 1.1 & 4.0e-13 & 0.525 & -1.0 & 3.13 \\ 
$10^{-0.25}$ & 3.0e-15 & 0.638 & 0.7 & 1.3 & 3.1e-13 & $\infty$ & - & 2.22 \\ 
$10^{0.0}$ & 2.7e-15 & 1.262 & 0.7 & 1.1 & 2.2e-13 & $\infty$ & - & 3.27 \\ 
$10^{0.25}$ & 1.5e-15 & 1.357 & 0.7 & 1.1 & 1.4e-13 & $\infty$ & - & 2.59 \\ 
$10^{0.5}$ & 9.8e-16 & 2.369 & 1.0 & 1.0 & 8.8e-14 & $\infty$ & - & 2.37 \\ \hline
\end{tabular}
\end{center}
\end{table}

\twocolumn

\appendix
\section{DIFFERENTIAL CROSS-SECTIONS FOR COULOMB SCATTERING}
\label{DIFFERENTIAL CROSS-SECTIONS FOR COULOMB SCATTERING}

\subsection{Electron-proton scattering: Rutherford cross-section}

Differential cross section is given by 
\begin{equation}
{d \sigma \over d \Omega}(\beta,\alpha)_{\rm Rutherford} 
= {3 \sigma_{\rm T} \over 32 \pi \gamma^2 \beta^4}
\left( 1+\frac{2 \gamma m_{\rm e}}{m_{\rm p}} \right)
{1 \over \sin^4 (\alpha/2)}, 
\end{equation}
\citep{ste83b},
where $\beta$ and $\gamma$ is the velocity of electron and its Lorentz factor
in the CM frame and $\sigma_{\rm T}$ is the Thomson cross section.
From equations (\ref{de}) and (\ref{dp}),
we need to integrate $2 \sin^2 (\alpha/2) {d \sigma \over d \Omega}$
over the scattering angles to obtain the energy and momentum transfer.
Therefore, we define the effective cross-section as 
\begin{equation}
\sigma_{\rm eff}  \equiv  \int^\pi_{\alpha_{\rm min}}
(1-\cos \alpha) {d \sigma \over d \Omega} 2 \pi \sin \alpha d \alpha 
. 
\end{equation}
Here $\alpha_{\rm min}$ is the lower limit of scattering angle.
Then we obtain 
\begin{equation}
\sigma_{\rm eff,Rutherford} =  {3 \sigma_T \over 2 \gamma^2 \beta^4}
\left( 1+\frac{2 \gamma m_{\rm e}}{m_{\rm p}} \right) \ln{\Lambda}, 
\end{equation}
where
\begin{equation}
\ln \Lambda \equiv \ln \left\{ {1 \over \sin(\alpha_{\rm min}/2)} \right\},
\end{equation}
is the Coulomb logarithm and usually evaluated about $10 \sim 20$.

\subsection{Electron-electron scattering: M{\o}ller cross-section}

The differential cross-section of e${^-}$-e${^-}$ scattering  \citep{jau80}
is given by 
\begin{equation}
{d \sigma \over d \Omega} (\beta,\alpha)_{\rm M{\o}ller}
={3 \sigma_{\rm T} \over 8 \pi \gamma^2 (\gamma^2 -1)^2 }
\left[ {(2 \gamma^2-1)^2 \over \sin^4 \alpha} -
 {2 \gamma^4 - \gamma^2- {1 \over 4} \over \sin^2 \alpha}
 + {(\gamma^2-1)^2 \over 4} \right]. 
\end{equation}
Since the particles are not identical, $\alpha$ is from $\alpha_{\rm min}$
to $\pi/2$.
In the same way as Rutherford scattering,
we obtain the effective cross-section with $\Lambda \gg 1$ as
\begin{eqnarray}
\sigma_{\rm eff,M{\o}ller} & = & {3 \sigma_T \over 16 \gamma^6 \beta^4} \left\{
 (2 \gamma^2-1)^2 \left[ 2 \ln\Lambda +1\right] 
-4 \left( 2 \gamma^4 - \gamma^2 -{1 \over 4} \right) \ln{2}
+\frac{\gamma^4 \beta^4}{2}
 \right\}.
\end{eqnarray}

\subsection{Electron-positron scattering: Bhabha cross-section}

The differential cross-section of e${^-}$-e${^+}$ scattering  \citep{jau80}
is given by 
\begin{eqnarray}
{d \sigma \over d \Omega} (\beta,\alpha)_{\rm Bhabha}
& = & {3 \sigma_{\rm T} \over 128 \pi \gamma^2} \cdot \left\{
{1 \over [\gamma \beta \sin (\alpha/2)]^4} \left[ 1+ \left( 2 \gamma \beta \cos{\alpha \over 2} \right)^2 + 2 (\gamma\beta)^4 \left( 1+\cos^4 {\alpha \over 2} \right) \right]  \right.  \nonumber  \\
& &  \hspace{1cm}  
 - {1 \over [\gamma^2 \beta \sin(\alpha/2)]^2} \left[3+ 2 \left( 2 \gamma\beta \cos {\alpha \over2} \right)^2 + {1 \over 4} \left( 2 \gamma\beta \cos {\alpha \over 2} \right)^4  \right]  \nonumber  \\
& & \hspace{1cm} 
\left. + {1 \over \gamma^4} \left[ 3+4(\gamma\beta)^2 + (\gamma\beta)^4 (1+ \cos^2 \alpha) \right]
\right\}
. 
\end{eqnarray}
The effective cross-section is given by 
\begin{eqnarray}
\sigma_{\rm eff,Bhabha} & = & {3 \sigma_T \over 16} \left[ 
{2 (2 \gamma^2-1)^2 \over \gamma^6 \beta^4} \ln\Lambda
- {4 \gamma^2 +3 \over \gamma^6 \beta^2}
- {6 \gamma^4 +4 \gamma^2 -1 \over 2 \gamma^6 }
- {4 \beta^2 - 2 \beta^4 \over 3 \gamma^2}
 \right]
. 
\end{eqnarray}

\end{document}